\documentclass[journal]{IEEEtran}
\usepackage{amsmath,amsfonts}
\usepackage{array}
\usepackage[caption=false,font=normalsize,labelfont=sf,textfont=sf]{subfig}
\usepackage{textcomp}
\usepackage{stfloats}
\usepackage{url}
\usepackage{verbatim}
\usepackage{graphicx}
\usepackage{cite}
\usepackage{calligra}
\usepackage{algorithm}
\usepackage{algorithmicx}
\usepackage{algpseudocode}
\usepackage{stfloats}
\usepackage{xcolor}
\usepackage{url}
\usepackage[hidelinks]{hyperref}

\graphicspath{{./figures/}}
\begin{document}

\title{Static Deep Q-learning for Green Downlink C-RAN}

\author{Yuchao~Chang,~Hongli~Wang,~Wen~Chen,~\IEEEmembership{Senior~Member,~IEEE},\\
~Yonghui~Li,~\IEEEmembership{Fellow,~IEEE},~and~Naofal~Al-Dhahir,~\IEEEmembership{Fellow,~IEEE}
\thanks{Y. Chang, H. Wang, and W. Chen are with the Department of Electronic Engineering, Shanghai Jiao Tong University, Shanghai 200240, China (e-mail: yuchaoc@mail.ustc.edu.cn; hongliwang@sjtu.edu.cn; wenchen@sjtu.edu.cn).}
\thanks{Y. Li is with the School of Electrical and
Information Engineering, The University of Sydney, Sydney, NSW 2006, Australia (email: yonghui.li@sydney.edu.au).}
\thanks{N. Al-Dhahir are with Department of Electrical and Computer Engineering, The University of Texas at Dallas, Richardson, TX 75080, USA (email: aldhahir@utdallas.edu).}
\thanks{(\emph{Corresponding author: Wen Chen.})}}
\markboth{Journal of \LaTeX\ Class Files,~Vol.~14, No.~8, August~2021}%
{Shell \MakeLowercase{\textit{et al.}}: Collaborative Multi-BS Power Management for Dense Radio Access Network using Deep Reinforcement Learning}

\maketitle

\begin{abstract}
Power saving is a main pillar in the operation of wireless communication systems. In this paper, we investigate cloud radio access network (C-RAN) capability to reduce power consumption based on the user equipment (UE) requirement. Aiming to save the long-term C-RAN energy consumption, an optimization problem is formulated to manage the downlink power without degrading the UE requirement by designing the power offset parameter. Considering stochastic traffic arrivals at UEs, we first formulate the problem as a Markov decision process (MDP) and then set up a dual objective optimization problem in terms of the downlink throughput and power. To solve this optimization problem, we develop a novel static deep Q-learning (SDQL) algorithm to maximize the downlink throughput and minimize the downlink power. In our proposed algorithm, we design multi-Q-tables to simultaneously optimize power reductions of activated RRHs by assigning one Q-table for each UE. To maximize the accumulative reward in terms of the downlink throughput loss and power reduction, our proposed algorithm performs power reductions of activated RRHs through continuous environmental interactions. Simulation results\footnote{Code can be accessed on website: \url{https://github.com/yuchaoch/project/tree/main}} show that our proposed algorithm enjoys a superior average power reduction compared to the activation and sleep schemes, and enjoys a low computational complexity.
\end{abstract}


\section{Introduction}
The projected increasing requirements of wireless traffic and green communications pose significant challenges for 5G and beyond wireless communication design\cite{Gvozdenovic2023IoT, Chen2023JSAC, Chang2017Sensors}. Cloud radio access network (C-RAN) was proposed to improve the network resource utilization by enhancing the network performance and reducing power consumption \cite{SaxenaN2016JSAC, pompili2016elastic, peng2016recent}. Different from the traditional radio base station architecture, C-RAN has the potential to reap valuable economic benefits for mobile networks operators (MNOs). 
C-RAN has shown great potential in providing unprecedented scalable, flexible, and efficient infrastructure provisioning and management in 5G and beyond wireless communication systems. C-RAN is composed of two parts: low-cost remote radio heads (RRHs) and powerful computing central baseband unit (BBU), where RRH and BBU communicate with each other through high-speed optical fiber transmission links \cite{wu2012green}. In the C-RAN architecture, RRH performs basic signal processing functions, and BBU aggregates powerful computing capability to centralized baseband processing and management, which makes it possible to guarantee low latency and high-rate data transmission \cite{marotta2017characterizing}. By dynamically adjusting resource allocation, the C-RAN architecture can not only significantly improve the network performance, but also reduce unnecessary energy consumption \cite{ChangY2018LCOMM}. Moreover, RRHs can be densely deployed around user equipments (UEs) at low operating costs, which contributes to significantly reducing the transmission power \cite{ICC2021Yue}.
\par
Recently, global carbon emissions have continued to rise, resulting in more harsh global climate issues. The report in \cite{ICT2021The} pointed out that the information and communication technology (ICT) share of global greenhouse gas (GHG) emissions was estimated to be 1.8-2.8\% of the global GHG emissions in 2021. Meanwhile, the ICT industry accounts for nearly 20\% of the total electricity consumption and keeps an annual growth rate of more than 6\% \cite{mao2021ai}. Li \emph{et al} pointed out in \cite{TACT2014_RLi} that over 80\% of the power consumption took place in the radio access networks (RAN). Several research efforts have been carried out to address green communications \cite{mahapatra2015energy, zhang2016fundamental, BBDai2016JSAC, sigwele2020energy, TCOMM.2021.3091133, Mao2022COMST319, TCM2021Teng, RTao2019TWC, qin2020green, TVT.2017.2719404, TVT2016Bousia}. In \cite{mahapatra2015energy}, an extensive survey of energy efficiency techniques was presented. 
Zhang \emph{et al} analyzed four basic network tradeoffs: spectrum efficiency versus energy efficiency, deployment efficiency versus energy efficiency, delay versus power, and bandwidth versus power \cite{zhang2016fundamental}. Dai \emph{et al} developed a downlink C-RAN architecture to minimize the network power consumption \cite{BBDai2016JSAC}. Sigwele \emph{et al} studied the pico base stations switching-off strategy to maximize the energy efficiency in the  downlink C-RAN \cite{sigwele2020energy}. Rate splitting multiple access (RSMA) was developed to enhance power management in RAN \cite{TCOMM.2021.3091133}. A comprehensive tutorial on RSMA was provided in \cite{Mao2022COMST319} and analyzed issue of energy efficiency. 
Teng \emph{et al} in \cite{TCM2021Teng} studied traffic uncertainty for the joint optimization of base station (BS) activation and user association to mitigate interference and balance traffic loads. 
Tao \emph{et al} in \cite{RTao2019TWC} presented a novel sleeping mechanism for small cells to decrease the energy consumption of heterogeneous networks. 
A resource-on-demand energy scheduling strategy for multiple cooperating radio access technologies was proposed to tackle the varying network energy efficiency demand in wireless networks \cite{qin2020green}. 
Oikonomakou \emph{et al} proposed a scheme of switching off part of the small cells to enhance the energy efficiency of HetNets \cite{TVT.2017.2719404}. 
To improve energy savings, a multi-objective auction-based switching-off scheme in heterogeneous networks was developed \cite{TVT2016Bousia}.
Although these algorithms improved power savings, there is still huge potential for power conservation in 5G and beyond wireless communication systems, especially in accurate power management.
\par
In a 5G and beyond wireless communication system, there is no doubt that the volume of network infrastructure and the number of connected terminals will keep increasing exponentially, which not only makes the wireless networks more complex, but also results in surging energy consumption. For the complex ever-changing wireless network, ruled-based traditional optimization algorithms have shown their limitations, but artificial intelligence (AI) techniques can provide context-aware information transmissions and personal-customized services, as well as realize automatic network management \cite{chang2017distributed, letaief2019roadmap, YChangACCESS2019}. Hence, wireless network optimization techniques have moved from the rule-based to AI-based mode of operation. In C-RAN, aiming to adapt to the dynamically changing environment, flexible and efficient power management should be carefully designed. 
In \cite{gu2021knowledge}, the authors developed a knowledge-assisted deep reinforcement learning (DRL) algorithm to design wireless schedulers with time-sensitive traffic, which aimed to improve quality-of-service in the fifth-generation (5G) cellular networks. To address the significant training costs and convergence issues, Xu \emph{et al} designed a DRL-based scheduling strategy by adopting the same policy network to allocate every resource block group (RBG) \cite{xu2020buffer, wang2019deep}. 
Reinforcement learning (RL), more specifically Q-learning, shows powerful capabilities to solve complicated decision-making problems in dynamically varying environment \cite{Book_RL}. Q-learning has been applied to a wide range of wireless communication scenarios and achieved impressive results \cite{zhang2022q, ICCCS2021Ding, JCC.2021.08.016, JIOT.2021.3098331, TNSM.2023.3245544, OJCOMS.2022.3219014}. Zhang \emph{et al} developed a Q-learning framework to realize intelligent routing to maximize the utility for cognitive unmanned aerial vehicle swarm \cite{zhang2022q}. To tackle the problem of local optimal solution and slow convergence speed in micro-nano satellite networks, an improved quantum ant colony Quality of Service (QoS) routing algorithm using Q-learning was proposed in \cite{ICCCS2021Ding}. In \cite{JCC.2021.08.016}, Chen \emph{et al} proposed a Q-learning-based multi-hop cooperative routing protocol to improve the quality of routing for underwater acoustic sensor networks. To achieve a realistic and comprehensive perspective of the decision process for customized service provision, the work in \cite{JIOT.2021.3098331} proposed a virtualized resource orchestration strategy in RAN of Internet of Things (IoT). An optimization problem was formulated to minimize offloading, fronthaul routing, and computation delays in Open RAN. To solve this NP-hard problem, the work in \cite{TNSM.2023.3245544} used deep Q-learning assisted by federated learning with a reward function that reduces the cost of delay. Al-hammadi \emph{et al} in \cite{OJCOMS.2022.3219014} formulated a joint power allocation and Light Emitting Diode (LED) transmission angle tuning optimization problem to maximize the average sum rate and the average energy efficiency.  As a simple RL algorithm, Q-learning has great potential to tackle wireless communication problems through the continuous environmental interaction.
\par
Deep learning refers to the learning models that involve a greater amount of composition of learned functions \cite{lecun2015deep}. Mnih \emph{et al} developed a reinforcement learning agent called deep Q-network (DQN) that combined Q-learning with a multiple-layer convolutional artificial neural network (ANN) to process spatial arrays of data such as images. Similarly, we develop a static deep Q-learning (SDQL) algorithm to maximize power saving without degrading the UE requirement in C-RAN. The basic idea of "deep" in SDQL refers to multiple Q-tables (denoted henceforth by deep Q-table) by improving the Q-learning algorithm. In the deep Q-table, all Q-tables are interdependent with each other and have identical state space and action space. When conducting the continuous optimization iterations, our proposed deep Q-table framework and its state and action space are free from changes, while rewards will be updated synchronously based on state transitions and taking actions. Hence, our proposed Q-table is deep and static, and our proposed Q-learning algorithm is named static deep Q-learning algorithm. The main contributions of this work are summarized as follows: 
\begin{enumerate}
\item We propose a novel static deep Q-learning algorithm to maximize the network-wide downlink throughput and power saving without degrading the UE throughput requirement. Based on the unpredictable differentiated traffic demands from UEs, each UE has its own throughput requirement. BBU simultaneously optimizes the downlink power of activated RRHs while satisfying UEs' throughput requirements.
\item To achieve precise downlink power reduction, we define a power offset for RRHs based on the UE throughput requirement. The UE throughput requirement determine whether the power offset is a positive or negative value, indicating whether we should perform power reduction for its associated RRH or not. Power reduction for RRH, in turn, affects the power offset value through interference reduction.
\item We formulate a dual objective optimization problem to simultaneously maximize the network-wide downlink throughput and power reduction. The UE throughput requirement and downlink power are defined as constraints. The downlink power variation for activated RRHs determines the desired and interference signals at UEs, making the two optimization objectives interdependent.
\item To tackle the dual objectives' optimization problem, we propose a static deep Q-learning algorithm to maximize the accumulative reward. The Reference Signal Receiving Power (RSRP) at UEs and power reduction at RRHs are, respectively, defined as the state and action space, and the additive value of power reduction and throughput loss is defined as the reward to perform the dual objectives' optimization. Moreover, our proposed algorithm continuously performs power reduction through continuous environmental interactions. 
\item Simulation results in diverse scenarios show that our proposed algorithm enjoys a superior power reduction compared to the activation and sleep schemes, and also has a low computational complexity. The network-wide throughput suffers some loss due to power reduction, but does not degrade UEs' throughput requirements. Meanwhile, power reductions make unsatisfied UEs suffer in decreasing interference, which makes some unsatisfied UEs become satisfied UEs.
\end{enumerate}
\par
This paper is structured as follows: Section \ref{System_Model} describes the system model, where the radio communication model, desired throughput analysis, and problem formulation are described. Section \ref{QL_Problem_solution} presents the Q-learning aided problem solution, in which we model the downlink power management problem as a Markov decision process (MDP), and extend Q-learning into the static deep Q-learning to tackle the downlink power management in C-RAN. Section \ref{Numerical_Results} demonstrates the superiority of our proposed algorithm in terms of both power saving and UE satisfaction. We also analyzed the convergence and optimization iterations of SDQL. Section \ref{Conclu} concludes our work. Moreover, for ease of reference, \textbf{Table} \ref{tab:No01} lists the main acronyms. To the best of our knowledge, this paper is the first which proposes to extend Q-learning into the static deep Q-learning. 
\begin{table}[!t]
\caption{Summary of main acronyms}
\centering
\label{tab:No01}
\begin{tabular}{|c||c|}
\hline
\textbf{Acronym} & \textbf{Meaning} \\
\hline
  5G & Fifth generation \\
  \hline
  RAN & Radio access network \\
  \hline
  C-RAN & Cloud radio access network \\
  \hline
  MNO &  Mobile network operator \\
  \hline
  UE & User equipment \\
  \hline
  BS & Base station \\
  \hline
  MDP &  Markov decision process \\
  \hline
  RSRP & Reference Signal Receiving Power   \\
  \hline
  SDQL& Static deep Q-learning \\
  \hline
  RRH & Remote radio head \\
  \hline
  BBU & Baseband unit \\
  \hline
  ICT & Information and communications technology \\
  \hline
  GHG & Global greenhouse gas \\
  \hline
  RSMA & Rate splitting multiple access \\
  \hline
  RL & Reinforcement learning  \\
  \hline
  DQN & Deep Q-network \\
  \hline
  IoT & Internet of Things \\
  \hline
  LED & Light Emitting Diode \\
  \hline
  QoS & Quality of Service \\
  \hline
  ANN & Artificial neural network \\
  \hline
  SINR & Signal-to-interference-plus-noise-ratio \\
  \hline
  CDF & Cumulative distribution function   \\
  \hline
  KPI & Key performance indicator \\
  \hline
  AI & Artificial intelligence \\
\hline
\end{tabular}
\end{table}
\begin{table*}[!t]
\caption{Summary of main notation}
\centering
\label{tab:No02}
\begin{tabular}{|c|c|c|c|c|}
\hline
\multicolumn{2}{|c|}{\textbf{Problem formulation}} &  & \multicolumn{2}{c|}{\textbf{Static deep Q-learning}}\\
\hline
\textbf{Notation} & \textbf{Description} & & \textbf{Notation} & \textbf{Description}\\
  \hline
  ${\mathcal B}$  & Set of RRHs & & $h_{b,u}^{n}$ & The available $n^{th}$ state for UE $u$ \\
  \hline
  ${\mathcal U}$  & Set of UEs & &  ${{\mathcal H}_{b,u}}$ & The RSRP set for UE $u$ \\
  \hline
  ${y_{b,u}}$ &  The received signal for UE $u$ & & $s_{\mathcal U}^{(k)}$ & The $k^{th}$ C-RAN state \\
  \hline
  ${h_{b,u}}{s_{u}}$ &  The desired signal for UE $u$ & & ${\mathcal S}$ & The state space \\
  \hline
  $n_0$ & The additive white Gaussian Noise & & $K$ & The number of elements in ${\mathcal S}$ \\
  \hline
  $P_{b,u}$ & The downlink power & & ${\bf{P}^{\Delta}}$ & The unified action set \\
  \hline
  $H_{b,u}$ & The channel gain & & ${\mathcal P}_{b,u}^{\Delta,(k)}$ & The available action set for RRH $b$ \\
  \hline
  $P_{\max }$ & The maximum transmit power & & $\varepsilon$ & The exploration rate \\
  \hline
  ${\gamma _{b,{u}}}$ & SINR for UE $u$ & & $\Delta P_{b,{u}}^{(k)}$ & The chosen power reduction for RRH $b$ \\
  \hline
  $\sigma ^2$ & The noise power & & $Q\left( {h_{b,{u}}^{\left( n \right)},\Delta P_{b,u}^{(k)} } \right)$ & The Q-value for UE $u$ \\
  \hline
  $R_{b,u}$ & The downlink throughput for RRH $b$ & & $a_{\mathcal B}^{(k)}$ & The $k^{th}$ action set \\ 
  \hline
  $W$ & The bandwidth & & ${\mathcal A}$ & The action space \\
  \hline
  ${\widetilde \gamma _{b,u}}$ & The desired SINR for UE $u$  & & $\Delta {R_{b,u}^{(k)}}$ & The throughput loss for UE $u$ \\
  \hline
  ${\widetilde R_{b,u}}$ & The desired throughput for UE $u$  & & $q_{b,u}^{\left( k \right)}$ & The immediate reward for UE $u$  \\
  \hline
  ${\widetilde P_{b,u}}$ &  The desired downlink power for UE $u$ & & $w_0, w_1$ & The weights for reward \\
  \hline
  $H_{TX}$ & the signal gain constants at transmitter & & $\alpha$ & The learning rate \\
  \hline
  $c$ & The speed of light & & $\lambda$ & The discount rate \\
  \hline
  $f_c$ & The center frequency &  & ${\mathcal Q}\left( {{{\mathcal H}_{b,u}},{\bf{P}}^\delta } \right)$ & Q-table for UE $u$ \\
  \hline
   & &  & $\Delta {\overline P _{{\rm{Reduction}}}}$ & The average power reduction \\
  \hline
   & &  & $\Delta {\overline P _{{\rm{Offset}}}}$ & The average power offset \\
  \hline
   & &  & $\Delta {\overline P _{{\rm{Interference \  reduction}}}}$ & The average interference reduction \\
  \hline
   & &  & $\Delta {\overline P _{{\rm{Interference}}}}$ & The average interference \\
  \hline
\end{tabular}
\end{table*}
\section{System Model}\label{System_Model}
\begin{figure}[htbp]
\centerline{\includegraphics[width=3.25in]{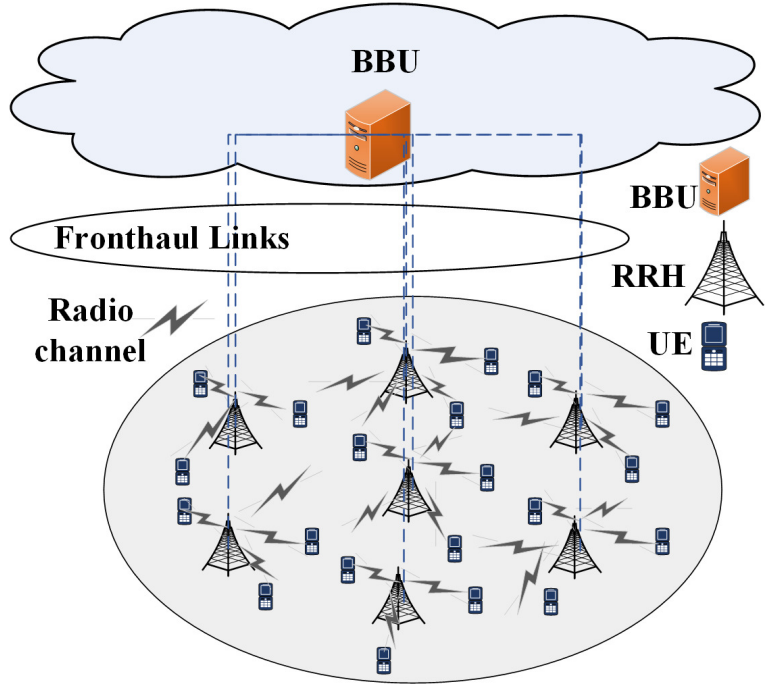}}
\caption{C-RAN architecture deploying BBU, RRHs, and UEs.}
\label{Fig1_Net_Topo}
\end{figure}

We analyze a downlink C-RAN system, including one BBU and $B$ RRHs. Each RRH is equipped with $N$ transmit antennas and serves UEs in the same time-frequency resource \cite{ICC2021Yue}, as depicted in \textbf{Fig. }\ref{Fig1_Net_Topo}. The set of RRHs and UEs are, respectively, denoted by ${\mathcal B} = \left\{ {0, \ldots ,b, \ldots ,B-1} \right\}$ and ${\mathcal U} = \left\{ {0, \ldots ,u, \ldots ,U-1} \right\}$. The time-step structure is adopted in our C-RAN system, where every time-step is defined as a learning period and regulated by 1 msec \cite{TGCN2023Chang}. RRHs are pre-deployed with fixed locations by the operator, while each UE can wander to different locations and report its location to its serving RRH in a fixed time interval such as 2 seconds. Since one learning period is a very short time interval, we ignore the location change of UEs within one learning period. We assume that  each RRH can serve numerous UEs and each UE can receive signals from multiple RRHs  at the same time. Each UE chooses the RRH with the maximal RSRP as its serving RRH when it has many candidate RRHs \cite{WCLopez2019}. 
We adopt the link budget calculation to explore the optimization problem of downlink power management to simplify the channel propagation model and lower the estimation overhead \cite{WCLopez2019}. Based on the link budget calculation, the strength of the received signal at UE is measured by RSRP, which is useful to estimate whether RSRP for each UE is sufficient or not \cite{RoseHu2013Book}. In the long-term C-RAN system, the UEs' desired throughput is random and diverse, resulting in network interference fluctuating over time. For ease of reference, \textbf{Table} \ref{tab:No02} lists the main notations of the system model.
\subsection{Radio Communication Model}\label{RCM}
In the studied C-RAN system, assuming RRH $b$ is chosen as the serving RRH for UE $u$ based on the maximal RSRP policy \cite{WCLopez2019}. 
The received signal for UE $u$ \cite{ICC2021Yue} can be expressed as follows
\begin{equation}\label{y_k_desired_signal}
    \begin{gathered}
      {y_{b,u}} = {h_{b,u}}{s_{u}}  + \sum\limits_{b^{'} \in {\mathcal B}\backslash b} {\sum\limits_{v \in {\mathcal U}\backslash u} {{h_{b^{'},v}}{s_v}} } + {n_0}, 
    \end{gathered}
\end{equation}
where the first term ${h_{b,u}}{s_{u}}$ is the desired signal, the second term $\sum\limits_{b^{'} \in {\mathcal B}} {\sum\limits_{v \in {\mathcal U}\backslash u} {{h_{b^{'},v}}{s_v}} }$ is the interference signal, and the third term $n_0$ denotes the additive white Gaussian Noise. The term $h_{b,u}$ is denoted by
\begin{equation}\label{h_b_gain_signal}
    \begin{gathered}
      h_{b,u}={P_{b,u}}{H_{b,u}},
    \end{gathered}
\end{equation}
where $H_{b,u}$ is the channel gain that accounts for the total transmitter gain, path loss, and total receiver gain from RRH $b$ to UE $u$ \cite{TGCN2023Chang}. $H_{b,u}$ is defined by 
\begin{equation}\label{H_b_u_gain}
    {H_{b,u}} = {H_{TX}} \times \left( {\frac{c}{{4\pi  \times {f_c} \times {D_{b,u}}}}} \right),
\end{equation}
where $H_{TX}$ is the signal gain at the transmitter, $c$ is the speed of light, $f_c$ is the center frequency, and $D_{b,u}$ is the distance between RRH $b$ and UE $u$. 
$P_{b,u}$ is the downlink power level and is subject to a power constraint defined by ${P_{b,u}} \le {P_{\max }}$. All downlink powers for RRHs are subject to the power constraint. $P_{\max }$ is the maximum transmit power level and is identical for all RRHs. The power unit translation between "dBW" and "W" \cite{TGCN2023Chang} is denoted by
\begin{equation}\label{dBw_W_Unit}
    P_{b,u}\left[ {{\rm{in \  dBW}}} \right] = 10 \cdot {\log _{10}}\left( {\frac{{P_{b,u}\left[ {{\rm{in \  W}}} \right]}}{{1\left[ {{\rm{in \  W}}} \right]}}} \right).
\end{equation}
Hence, the signal-to-interference-plus-noise-ratio (SINR) for UE $u$ is given by
\begin{equation}\label{sinr}
    \begin{gathered}
      \begin{aligned}
        {\gamma _{b,{u}}} &= \frac{{{h_{b,u}}}}{{\sum\limits_{b^{'} \in {\mathcal B}\backslash b} {\sum\limits_{v \in {\mathcal U}\backslash u} {{h_{b^{'},v}}}  + {\sigma ^2}} }}\\
         &= \frac{{{P_{b,u}}{H_{b,u}}}}{{\sum\limits_{b^{'} \in {\mathcal B}\backslash b} {\sum\limits_{v \in {\mathcal U}\backslash u} {{P_{b^{'},v}}{H_{b^{'},v}}}  + {\sigma ^2}} }},
      \end{aligned}
    \end{gathered}
\end{equation}
where $\sigma ^2$ is the noise power. Consequently, the downlink throughput \cite{ICC2020Marzouk} is given by 
\begin{equation}\label{dl_data_rate}
    \begin{gathered}
      {R_{b,u}} = W{\log _2}\left( {1 + {\gamma _{b,u}}} \right),
    \end{gathered}
\end{equation}
where $W$ is the bandwidth.
\subsection{Desired Throughput Analysis}\label{QoS_analysis}
The network downlink throughput is one of the core key performance indicators (KPI) in C-RAN. The research in \cite{yifei2014application} provides conventional scenario classifications and throughput requirements for different scenarios. Based on the throughput requirement definition, all UEs can be dynamically divided into satisfied central UEs and unsatisfied weak UEs. For example, UE $u$ is served by RRH $b$ and has the desired throughput ${\widetilde R_{b,u}}$. UE $u$ is a satisfied central UE as ${R_{b,u}} \ge {\widetilde R_{b,u}}$, otherwise UE $u$ is an unsatisfied weak UE. 
Based on the definition in Equation (\ref{dl_data_rate}), the desired SINR for UE $u$ is denoted by
\begin{equation}\label{UE_desired_SINR}
    \begin{gathered}
      {\widetilde \gamma _{b,u}} = {2^{\frac{{{{\widetilde R}_{b,u}}}}{W}}} - 1.
    \end{gathered}
\end{equation}
Combining the standard SINR definition in Equation (\ref{sinr}) and the desired SINR definition in Equation (\ref{UE_desired_SINR}), the desired downlink power for RRH $b$ can be expressed as
\begin{equation}\label{RRH_desired_power}
    \begin{gathered}
      \begin{aligned}
        {\widetilde P_{b,u}} &= \frac{1}{{{H_{b,u}}}}{\widetilde \gamma _{b,u}}\left( {\sum\limits_{{b^{'}} \in {\mathcal B}\backslash b} {\sum\limits_{v \in {\mathcal U}\backslash u} {{P_{{b^{'},v}}}{H_{{b^{'}},v}}}  + {\sigma ^2}} } \right)\\
        &= \frac{{{2^{\frac{{{{\widetilde R}_{b,u}}}}{W}}} - 1}}{{{H_{b,u}}}}\left( {\sum\limits_{{b^{'}} \in {\mathcal B}\backslash b} {\sum\limits_{v \in {\mathcal U}\backslash u} {{P_{{b^{'},v}}}{H_{{b^{'}},v}}}  + {\sigma ^2}} } \right).
      \end{aligned}
    \end{gathered}
\end{equation}
Hence, the power offset for RRH $b$ can be calculated as follows
\begin{equation}\label{RRH_oft_power}
    \begin{gathered}\!\!\!
      \begin{aligned}
        P_{b,u}^{\delta} &= {P_{b,u}} - {\widetilde P_{b,u}}\\
         &=\!\!{P_{b,u}}\!-\!\frac{{{2^{\frac{{{{\widetilde R}_{b,u}}}}{W}}} - 1}}{{{H_{b,u}}}}\left( {\sum\limits_{{b^{'}} \in {\mathcal B}\backslash b} {\sum\limits_{v \in {\mathcal U}\backslash u} {{P_{{b^{'},v}}}{H_{{b^{'}},v}}}  + {\sigma ^2}} } \right)\!\!.
      \end{aligned}\!\!\!
    \end{gathered}
\end{equation}
Equation (\ref{RRH_oft_power}) indicates that the power offset $P_{b,u}^{\delta}$ for RRH $b$ is impacted by the downlink power of activated RRHs. Therefore, power offsets for activated RRH are interdependent with each other, forming a collaborative power management optimization problem. Power reductions for other activated RRHs are calculated based on their power offsets.
\subsection{Problem Formulation}\label{formulation_problem}
As is well known, MNOs do not only achieve economic benefits from maximizing the network-wide downlink throughput, but also reduce energy consumption costs from power reductions of RRHs. However, power reductions of RRHs possibly bring throughput loss to their associated UEs, which depends on the power reductions between the desired signal and the interference signal. Therefore, the downlink throughput and power involve a design tradeoff, which inspires us to formulate an optimization problem compromising the network-wide downlink throughput and power. 
Based on the diverse desired throughput from UEs, the network-wide downlink throughput maximization in C-RAN is defined by
\begin{equation}\label{maxi_throughput}
    \begin{gathered}
      \begin{array}{l}
        \left( { {\bf{P0}}} \right) \quad  \max \sum\limits_{b \in {\mathcal B}} {\sum\limits_{{u} \in {\mathcal U}} {{R_{b,{u}}}} } \\
        {\rm s.t.}: {P_{b,u}} \le {P_{\max }},\forall b \in {\mathcal B},\forall u \in {\mathcal U}.
      \end{array}
    \end{gathered}
\end{equation}
Problem $(\bf{P0})$ means that C-RAN tries to maximize the network-wide downlink throughput. 
\par
When satisfying UEs' throughput requirements, another objective is to minimize the downlink power for RRHs in C-RAN. Hence, the network-wide downlink power minimization is defined as follows
\begin{equation}\label{min_power}
    \begin{gathered}
      \begin{array}{l}
         \left( { {\bf{P1}}} \right) \quad  \min \sum\limits_{b \in {\mathcal B}} {\sum\limits_{{u} \in {\mathcal U}} {{P_{b,{u}}}} } \\
         {\rm s.t.}: {R_{b,u}} \ge {\widetilde R_{b,u}},\forall b \in {\mathcal B},\forall u \in {\mathcal U}.        
      \end{array}
    \end{gathered}
\end{equation}
Problem $(\bf{P1})$ means that C-RAN tries to minimize the downlink power consumption. Accordingly, to simultaneously satisfy the downlink throughput maximization and the downlink power minimization, we integrated Problem $(\bf{P0})$ and $(\bf{P1})$ to form a dual objective optimization problem as follows
\begin{equation}\label{problem_optimization}
    \begin{gathered}
      \begin{array}{l}
        \left( {{\bf{P2}}} \right){\rm{   }}\left\{ \max \sum\limits_{b \in {\mathcal B}} {\sum\limits_{{u} \in {\mathcal U}} {{R_{b,{u}}}} , \quad { \min \sum\limits_{b \in {\mathcal B}} {\sum\limits_{{u} \in {\mathcal U}} {{P_{b,{u}}}} }  } } \right\}\\
        {\rm s.t.}\left\{ \begin{array}{l}
        {R_{b,u}} \ge {\widetilde R_{b,u}},\forall b \in {\mathcal B},\forall u \in {\mathcal U},\\
        {P_{b,u}} \le {P_{\max }},\forall b \in {\mathcal B},\forall u \in {\mathcal U}. 
        \end{array} \right.
      \end{array}
    \end{gathered}
\end{equation}
In Problem $(\bf{P2})$, the two optimization objectives are interdependent. These two constraints are, respectively, the desired throughput requirements from UEs and the downlink power consumption from RRHs. How to tackle these two optimization objectives is a non-convex optimization problem as well as NP-hard.
\section{Q-LEARNING AIDED PROBLEM SOLUTION}\label{QL_Problem_solution}
The C-RAN system has made the complex optimization techniques of resource allocation move from rule-based to AI-based. RL, more specifically Q-learning, is a gradual AI learning process that constantly adapts to the environment. We propose a novel static deep Q-learning to tackle the optimization in Problem $(\bf{P2})$.
\par
We pursue an intelligent power optimization aided by Q-learning approach to maximize the downlink throughput and to minimize the downlink power without degrading UEs' requirements in C-RAN. Our proposed algorithm considers, respectively, BBU and RRHs as the learning agent and action executors. Integrating the proposed Deep Q-table, the learning process is modeled as an improved MDP through continuous environmental interactions. Our proposed algorithm is constructed by 6-tuple $({\mathcal S}, {\mathcal A}, {\mathcal R}, P( {h_{b,u}^{\left( k \right)},h_{b,u}^{\left( {k + 1} \right)}|\Delta P_{b,u}^{\left( k \right)}} ), \lambda, $ Deep Q-table$)$. Specifically, the MDP elements consist of the state space ${\mathcal S}$, the action space ${\mathcal A}$, the reward space ${\mathcal R}: {\mathcal S} \times {\mathcal A} \to {\mathbb R}$, and the state transition probability $P( {h_{b,u}^{\left( k \right)},h_{b,u}^{\left( {k + 1} \right)}|\Delta P_{b,u}^{\left( k \right)}} )$. The discount factor determines the effect of future rewards on the current action. The deep Q-table consists of multiple Q-tables. The elements in the 6-tuple are defined as follows. \\
\par
(1) \emph{State}: For each UE, the state set is defined based on its RSRP. For example, the RSRP set for UE $u$ is denoted by
\begin{equation}\label{UE_u_state}
    \begin{gathered}
      {{\mathcal H}_{b,u}}\!=\!\left\{\!\!\!\begin{array}{l}
        \left[ {h_{b,u}^{ \left({ - H} \right)}, \ldots ,h_{b,u}^{\left( 0 \right)}, \ldots ,h_{b,u}^{ \left(n \right)}, \ldots ,h_{b,u}^{\left( H \right)}} \right]\!\!,\!\!\\
        h_{b,u}^{\left( n \right)} = \left\lfloor {{h_{b,u}}} \right\rfloor , \quad h_{b,u}^{\left( n \right)} \in {\mathbb Z},\\
        \forall b \in {\mathcal B}, \quad \forall u \in {\mathcal U}, \quad H \in {\mathbb Z}, 
        \end{array} \right\}\!\!\!
    \end{gathered}
\end{equation}
where $H$ is an integer constant, while being subject to $\left| n \right| \le H$. The state observed by the agent is determined by a combination of RSRPs for all UEs. Thus, we define the $k^{th}$ C-RAN state as follows
\begin{equation}\label{time_t_state}
    \begin{gathered}
      s_{\mathcal U}^{(k)} = \left( {{h_{0,0}^{(k)},} \ldots ,{h_{b,{u}}^{(k)}}, \ldots ,{h_{B,{U}}^{(k)}}} \right), 
    \end{gathered}
\end{equation}
where ${h_{b,{u}}^{(k)}} \in {{\mathcal H}_{b,u}}$. All possible C-RAN states, similar to $s_{\mathcal U}^{(k)}$, form the state space as follows
\begin{equation}\label{state_space}
    \begin{gathered}
      {\mathcal S} = \left\{ {s_{\mathcal U}^{(0)}, \ldots ,s_{\mathcal U}^{(k)}, \ldots ,s_{\mathcal U}^{(K)}} \right\}, 
    \end{gathered}
\end{equation}
where $k \in {\mathbb N}^+$. The total number of elements in ${\mathcal S}$ is given by $K=\left| {{{\mathcal H}_{0,0}}} \right| \times  \cdots  \times \left| {{{\mathcal H}_{b,u}}} \right| \times  \cdots  \times \left| {{{\mathcal H}_{B,U}}} \right|$, and $\left| {{{\mathcal H}_{b,u}}} \right|$ is the number of elements in ${{\mathcal H}_{b,u}}$. 
\par
(2) \emph{Action}: In C-RAN, the agent calculates the possible power reductions for activated RRHs. Each RRH receives the power reduction from BBU through the fronthaul link, and propagates data to its serving UE based on the reduced downlink power. Given that the power offset for each RRH is different, we adopt the slicing window to define the unified action set, which is expressed as ${\bf{P}^{\Delta}} = \left[ {0, \ldots ,P_\upsilon ^{\delta}, \ldots ,P_\Upsilon ^{\delta}} \right]\left( {P_\upsilon ^{\delta} \in {{\mathbb N}^ + }} \right)$, where $\Upsilon(\Upsilon \in {\mathbb N}^+)$ is the length of the slicing window. The available action set for RRH $b$ is determined by the power offset value and the unified action set, and is denote by 
\begin{equation}\label{RRH_action_set}
    \begin{gathered}
      \begin{aligned}
        {\mathcal P}_{b,u}^{\Delta,(k)}  &= \left[ {0, \ldots ,P_{b,u}^\delta } \right] \cap {{\bf{P}}^\Delta }\\
         &= \left\{ \begin{array}{l}
        \left[ 0 \right], \qquad \qquad \qquad \  P_{b,u}^\delta  \le 0,\\
        \left[ {0, \ldots ,\left\lfloor {P_{b,u}^\delta } \right\rfloor } \right], \quad 0 < P_{b,u}^\delta  \le P_\Upsilon ^\delta ,\\
        \left[ {0, \ldots ,P_\Upsilon ^\delta } \right], \qquad \ P_{b,u}^\delta  > P_\Upsilon ^\delta .
        \end{array} \right.
        \end{aligned}
    \end{gathered}
\end{equation}
To seek a tradeoff between exploration and exploitation, we adopt the $\varepsilon-$greedy policy in the learning process with the exploration rate $\varepsilon \in \left( {0,1} \right)$. More specifically, conditioned on the current state ${h_{b,{u}}^{(k)}}$ for RRH $b$, the agent chooses power reduction $\Delta P_{b,{u}}^{(k)}$ for RRH $b$, i.e.,
\begin{equation}\label{action_explo}
    \begin{gathered}\!\!
      \begin{array}{l}
        \Delta P_{b,{u}}^{\left( k \right)} = 
        \left\{\!\!\!\begin{array}{l}
        {\rm{random\ action\ }}{P_{b,{u}}^{{'}\left( k \right)}},\qquad \quad  \varepsilon ,\\
        \arg \mathop {\max }\limits_{P_{b,u}^{'\left( k \right)} \in {\mathcal P}_{b,u}^{\Delta,(k)} } Q\left( {h_{b,u}^{\left( k \right)},P_{b,u}^{'\left( k \right)}} \right), 1 - \varepsilon ,
        \end{array} \right.
      \end{array}
    \end{gathered}
\end{equation}
where ${P_{b,{u}}^{{'}\left( k \right)}} \in {\mathcal P}_{b,u}^{\Delta,(k)}$. $Q\left( {h_{b,{u}}^{\left( k \right)},\Delta P_{b,u}^{(k)} } \right)$ is the Q-value associated with action $\Delta P_{b,{u}}^{(k)}$ taken by RRH $b$ under state ${h_{b,{u}}^{(k)}}$. The power offset $P_{b,u}^{\delta}$ is gradually shrinking as the number of learning episodes increases and approaches 0. Since the power offset $P_{b,u}^{\delta}$ shrinks the available action set ${\mathcal P}_{b,u}^{\Delta,(k)}$ as the number of learning episodes increases. The available action exploration space gradually shrinks with the increasing number of learning episodes, resulting in reducing the role of exploration mode.
Thus, we define the $k^{th}$ action set for activated RRHs as follows
\begin{equation}\label{time_t_action}
    \begin{gathered}
     a_{\mathcal B}^{(k)} = \left( {\Delta P_{0,{0}}^{(k)} , \ldots ,\Delta P_{b,{u}}^{(k)}, \ldots ,\Delta P_{B,{U}}^{(k)}} \right), 
    \end{gathered}
\end{equation}
where $\Delta P_{b,{u}}^{(k)} \in {\mathcal P}_{b,u}^{\Delta,(k)}$. All possible actions, similar to $a_{\mathcal B}^{(k)}$, form the action space as follows
\begin{equation}\label{action_space}
    \begin{gathered}
      {\mathcal A} = \left\{ {a_{\mathcal B}^{(0)}, \ldots ,a_{\mathcal B}^{(k)}, \ldots ,a_{\mathcal B}^{(K)}} \right\}. 
    \end{gathered}
\end{equation}
\par
(3) \emph{Reward}: The immediate reward for each RRH is the additive value of the downlink throughput loss and power reduction in response to the state transition from current state $s_{\mathcal U}^{(k)} \in {\mathcal S}$ to next state $s_{\mathcal U}^{(k+1)} \in {\mathcal S}$ by executing the action $\Delta P_{b,{u}}^{\left( k \right)}$. The throughput loss is denoted as 
\begin{equation}\label{delt_thr_power}
    \begin{gathered}
        \Delta {R_{b,u}^{(k)}} = R_{b,u}^{(k)} - {R_{b,u}^{(k+1)}},
    \end{gathered}
\end{equation}
where $R_{b,u}^{(k)}$ and $R_{b,u}^{(k+1)}$ are, respectively, the downlink throughputs for the current state $s_{\mathcal U}^{(k)} \in {\mathcal S}$ and next state $s_{\mathcal U}^{(k+1)} \in {\mathcal S}$. 
Accordingly, an immediate reward $q_{b,{u}}^{\left( k \right)} \in {\mathcal R}$ is expressed as follows
\begin{equation}\label{immediate_reward}
    \begin{gathered}
      q_{b,u}^{\left( k \right)} =  \left( {{w_0} \cdot \Delta P_{b,u}^{\left( k \right)} - {w_1} \cdot \Delta R_{b,u}^{\left( k \right)}} \right), 
    \end{gathered}
\end{equation}
where $w_0$ and $w_1$ are adjustable weighting factors, while $w_0 + w_1 = 1$, $w_0 \geq 0 $, and $w_1 \geq 0$.  
\begin{figure*}[b]
\hrulefill
 \begin{equation}\label{Q_value_update}
      Q\left( {h_{b,{u}}^{\left( k \right)},\Delta P_{b,u}^{\left( k \right)}} \right) \leftarrow Q\left( {h_{b,{u}}^{\left( k \right)},\Delta P_{b,u}^{\left( k \right)}} \right) + \alpha \left( {q_{b,{u}}^{\left( k \right)} + \lambda \mathop {\max }\limits_{^{P_{b,u}^{'\left( k \right)} \in {\mathcal P}_{b,u}^{\Delta,(k)}}} Q\left( {h_{b,{u}}^{\left( {k + 1} \right)}, P_{b,u}^{'\left( k \right)}} \right) - Q\left( {h_{b,{u}}^{\left( k \right)},\Delta P_{b,u}^{\left( k \right)}} \right)} \right).
  \end{equation}
  \begin{equation}\label{state_transition_probability}
    \begin{gathered}
     {P}\left( {h_{b,u}^{\left( k \right)},h_{b,u}^{\left( {k + 1} \right)}| \Delta P_{b,u}^{\left( k \right)}} \right) = \left( {\varepsilon  \cdot \frac{1}{{\left| {{\mathcal P}_{b,u}^{\Delta ,\left( k \right)}} \right|}}} \right)\left( {1 - \varepsilon } \right). 
    \end{gathered}
\end{equation}
 \begin{equation}\label{QT_UE_space}
      \begin{array}{l}
        {\mathcal Q}\left( {{{\mathcal H}_{0,{u_0}}},{\bf P}^\Delta } \right) \buildrel \Delta \over = 
        \begin{array}{*{20}{c}}
        {}&0& \ldots &{P_\upsilon ^\delta }& \ldots &{P_\Upsilon ^\delta }\\
        {h_{b,{u}}^{\left( { - H} \right)}}&{Q\left( {h_{b,{u}}^{\left( { - H} \right)},0} \right)}& \ldots &{Q\left( {h_{b,{u}}^{\left( { - H} \right)},P_\upsilon ^\delta } \right)}& \ldots &{Q\left( {h_{b,{u}}^{\left( { - H} \right)},P_\Upsilon ^\delta } \right)}\\
         \vdots & \vdots & \ddots & \vdots & \ddots & \vdots \\
        {h_{b,{u}}^{\left( 0 \right)}}&{Q\left( {h_{b,{u}}^{\left( 0 \right)},0} \right)}& \ldots &{Q\left( {h_{b,{u}}^{\left( 0 \right)},P_\upsilon ^\delta } \right)}& \ldots &{Q\left( {h_{b,{u}}^{\left( 0 \right)},P_\Upsilon ^\delta } \right)}\\
         \vdots & \vdots & \ddots & \vdots & \ddots & \vdots \\
        {h_{b,{u}}^{\left( n \right)}}&{Q\left( {h_{b,{u}}^{\left( n \right)},0} \right)}& \ldots &{Q\left( {h_{b,{u}}^{\left( n \right)},P_\upsilon ^\delta } \right)}& \ldots &{Q\left( {h_{b,{u}}^{\left( n \right)},P_\Upsilon ^\delta } \right)}\\
         \vdots & \vdots & \ddots & \vdots & \ddots & \vdots \\
        {h_{b,{u}}^{\left( H \right)}}&{Q\left( {h_{b,{u}}^{\left( H \right)},0} \right)}& \ldots &{Q\left( {h_{b,{u}}^{\left( H \right)},P_\upsilon ^\delta } \right)}& \ldots &{Q\left( {h_{b,{u}}^{\left( H \right)},P_\Upsilon ^\delta } \right)}
        \end{array}
      \end{array}.
  \end{equation}
\hrulefill
\end{figure*}
\begin{figure*}[htbp]
\centerline{\includegraphics[width=5.in]{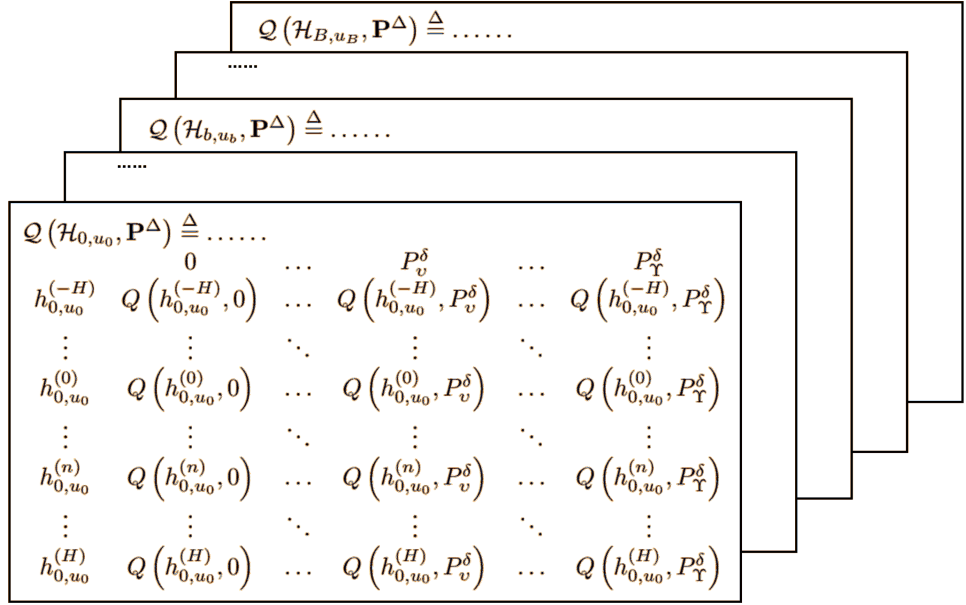}}
\caption{Deep Q-table.}
\label{Fig2_Deep_Q_table}
\end{figure*}
In $({\mathcal S}, {\mathcal A}, {\mathcal R}, P( {h_{b,u}^{\left( k \right)},h_{b,u}^{\left( {k + 1} \right)}|\Delta P_{b,u}^{\left( k \right)}} ), \lambda, $ Deep Q-table$)$, since $\lambda$ determines the effect of future rewards on the current action, the Q-value at each state can be updated in Equation (\ref{Q_value_update}), where $\alpha$ is the learning rate. Equation (\ref{Q_value_update}) shows the MDP state transition from the current state $h_{b,{u}}^{\left( k \right)}$ to the next state $h_{b,{u}}^{\left( k +1 \right)}$ by executing the action $\Delta P_{b,u}^{\left( k \right)}$.
\par
(4) \emph{State transition probability}: According to executing the action $\Delta P_{b,u}^{\left( k \right)}$, the state transition probability from the current state $h_{b,{u}}^{\left( k \right)}$ to the next state $h_{b,{u}}^{\left( k +1 \right)}$ is denoted in Equation (\ref{state_transition_probability}).
\par
(5) \emph{Discount factor}: The discount factor $\lambda$ determines the effect of future rewards on the current action. $\lambda$ is a hyperparameter and subject to $\lambda  \in \left[ {0,1} \right]$.
\par
(6) \emph{Deep Q-table}: Deep Q-table has multiple Q-tables and makes each Q-table correspond to one UE. For example, the Q-table for UE $u$ mainly consists of the RSRP set ${\mathcal H}_{b,u}$, action set ${\bf{P}}^\Delta$, and Q-values, which is expressed as ${\mathcal Q}\left( {{{\mathcal H}_{b,u}},{\bf{P}}^\Delta } \right)$ in Equation (\ref{QT_UE_space}). Q-tables for all UEs, similar to ${\mathcal Q}\left( {{{\mathcal H}_{b,u}},{\bf{P}}^\Delta } \right)$ in Equation (\ref{QT_UE_space}), form the deep Q-table as depicted in \textbf{Fig. }\ref{Fig2_Deep_Q_table}
\begin{algorithm}[t]
	\renewcommand{\algorithmicrequire}{\textbf{Input:}}
	\renewcommand{\algorithmicensure}{\textbf{Output:}}
	\caption{Static Deep Q-learning for Green in C-RAN}  
	\label{SDQL}
	\begin{algorithmic}[1]
		\Require $\mathcal B, {\mathcal U}, P_{\max}, W, \widetilde R_{b,u}, {\bf{P}^{\Delta}},\lambda, \alpha.$
        \State Initialize Q-table for each UE, $n_ \gamma = 0$, $n_ \Gamma = 10$.
        \State Obtain the C-RAN state defined in Equation (\ref{time_t_state})
        \For {$n=1$ to $N$ ($N$ is the iteration)}
           \For {$u=0$ to $U$}
               \State Calculate $P_{b,u}^{\delta}$ based on Equation (\ref{RRH_oft_power})
               \State Calculate ${\mathcal P}_{b,u}^{\Delta,(k)}$ based on Equation (\ref{RRH_action_set})
               \State Obtain $\Delta P_{b,{u}}^{\left( k \right)}$ based on Equation (\ref{action_explo})
               \State Calculate $R_{b,{u}}^{\left( k \right)}$ based on Equation (\ref{dl_data_rate})
               \State Calculate $q_{b,{u}}^{\left( k \right)}$ based on Eq. (\ref{immediate_reward})
               \State Update $Q\left( {h_{b,{u}}^{\left( k \right)},\Delta P_{b,u}^{\left( k \right)}} \right)$ based on Eq. (\ref{Q_value_update})
            \EndFor
            \State $\Gamma = \sum\limits_{b \in {\mathcal B}} {\sum\limits_{u \in {\mathcal U}} {q_{b,u}^{\left( k \right)}} }$
            \If{$\Gamma  =  = 0$}
                \If {$n_ \gamma == n_ \Gamma$}
                    \State Break
                \EndIf
            \EndIf
        \EndFor
	\end{algorithmic}
\end{algorithm}
\par
Our proposed static deep Q-learning algorithm is described in {\bf Algorithm \ref{SDQL}}. We describe the learning process of static deep Q-learning from lines 4 to 11, in which BBU optimizes the downlink power reductions for activated RRHs based on the above learning procedures. The termination condition for optimization is defined in lines 12 to 17, and indicates that the reward, being defined by the downlink throughput loss and power reduction, converges smoothly to 0.
\section{NUMERICAL RESULTS}\label{Numerical_Results}
In this section, extensive simulations are performed to evaluate our proposed algorithm. For convenience, the simulation parameters of RRHs and UEs are set to be identical. The main simulation parameters are shown in \textbf{Table }\ref{tab:No03}. 
\begin{table}[t]
\setlength{\abovecaptionskip}{-0.00cm}
\setlength{\belowcaptionskip}{-0.0cm}
\caption{Summary of simulation parameters}
\centering
\label{tab:No03}
\begin{tabular}{|c|c|c|c|c|}
\hline
\multicolumn{2}{|c|}{\textbf{Problem formulation}} &  & \multicolumn{2}{c|}{\textbf{Static deep Q-learning}}\\
\hline
\textbf{Parameter} & \textbf{Value} & & \textbf{Parameter} & \textbf{Value}\\
  \hline
  $P_{max}$  & 15.2dBW & & $\lambda$ & 0.9 \\
  \hline
  $n_0$  & -125dBW & &  $\varepsilon$ & 0.1 \\
  \hline
  $W$  & 10MHz & &  $\alpha$ & 0.1 \\
  \hline
  $H_{TX}$ & 17.5dBi & & $N$ & 100 \\
  \hline
  $c$ & $3\times10^{8}$ m/s & &  & \\
  \hline
  $f_c$ & $1.8\times10^{9}$ & &  & \\
  \hline
\end{tabular}
\end{table}
To better analyze the variations of throughput loss and power reduction, our proposed algorithm is compared with the activation and sleep schemes in \cite{TCM2021Teng, RTao2019TWC}. The work in \cite{TCM2021Teng} formulated the BS activation and user association problem as a mixed integer non-linear programming problem. Although the activation algorithm was able to ameliorate power efficiency by suppressing severe interference, it had to face the significant time delay and congestion risk due to the arrival of burst traffic \cite{alzubaidi2022interference}. In \cite{RTao2019TWC}, a two-tier HetNet with both the macro cell BS and small cell BS handling data plane was considered to achieve energy reduction by using the sleep mechanism. Meanwhile, the enhanced inter-cell interference coordination techniques cause a slight QoS degradation. We perform 1000 trials per simulation instance and average the results or show cumulative distribution function (CDF) results. 
\begin{figure*}[t]
  \centering{\includegraphics[width=7.15in]{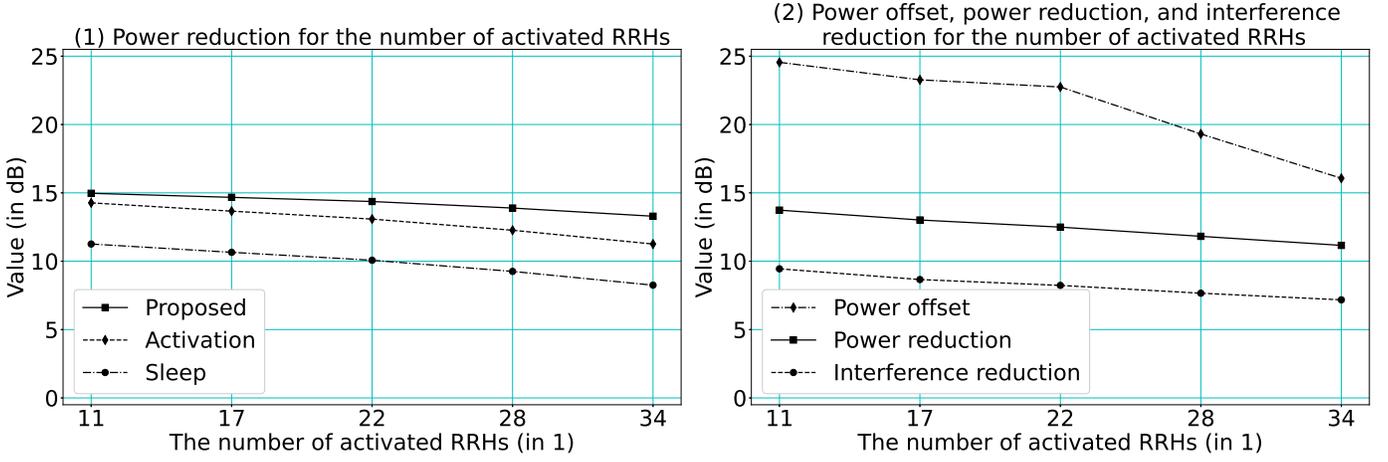}}
  \caption{Power reduction analysis on different number of activated RRHs.}\label{3Power_reduction_UE}
\end{figure*}
\begin{figure*}[t]
  \centering{\includegraphics[width=7.15in]{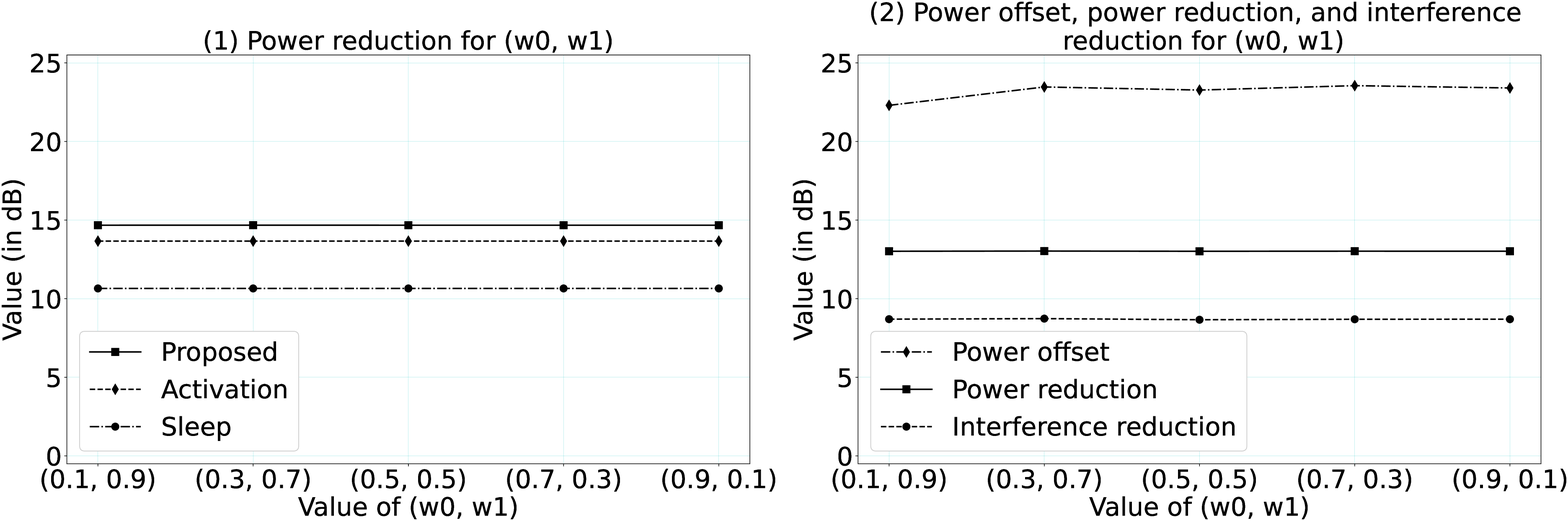}}
  \caption{Power reduction analysis on diverse values of $(w_0, w_1)$.}\label{4Power_reduction_weight}
\end{figure*}
\begin{figure}[t]
  \centering{\includegraphics[width=3.5 in]{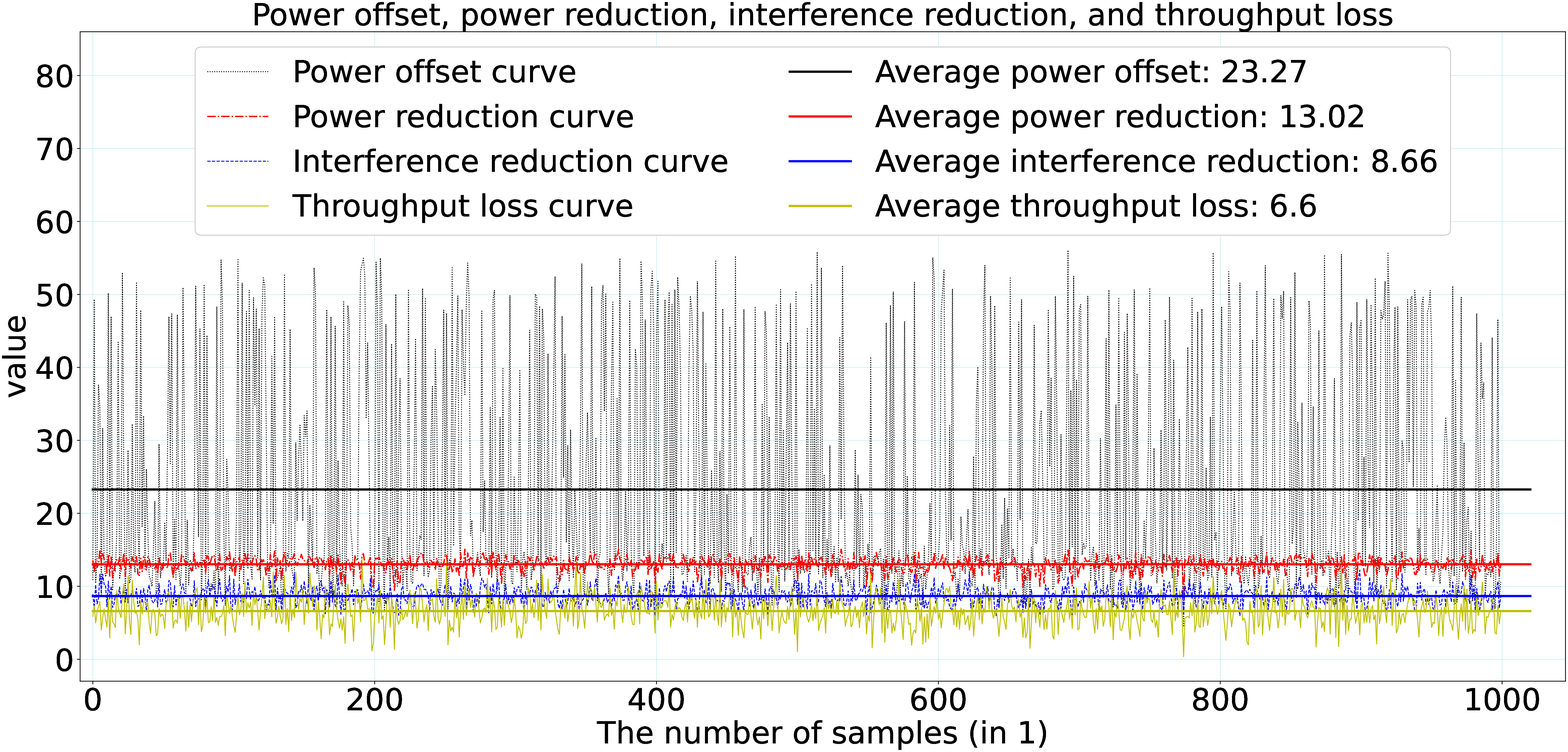}}
  \caption{Power offset vs power reduction vs interference reduction vs throughput loss.}\label{5RSRPOft_action_ITF_analysis}
\end{figure}
\subsection{Analysis for Power Reduction}\label{analysis_power_reduction}
In \textbf{Fig. }\ref{3Power_reduction_UE}, power reduction is analyzed based on the different numbers of activated RRHs. In particular, \textbf{Fig. }\ref{3Power_reduction_UE}-(1) shows the downward trend of the average power reduction for our proposed algorithm, activation algorithm, and sleep algorithm with the increasing number of activated RRHs in the C-RAN system. Meanwhile, the average power reduction for the C-RAN system is denoted as
\begin{equation}\label{avg_power}
    \begin{gathered}\!\!
      \begin{array}{l}
        \Delta {\overline P _{{\rm{Reduction}}}} = 
        10 \cdot {\log _{10}}\left( {\frac{{\sum\limits_{b = 0}^{B - 1} {\left( {{{10}^{\frac{{{P_{\max }}}}{{10}}}} - {{10}^{\frac{{{P_{_{b,u}}}}}{{10}}}}} \right)} }}{B}} \right)
      \end{array}\!\!.
    \end{gathered}
\end{equation}
For the C-RAN system with certain RRH set ${\mathcal B} = \left\{ {0, \ldots ,b, \ldots ,B-1} \right\}$, it is intuitive that these three algorithms exhibit downward trends in  power reduction with increasing number of activated RRHs based on Equation (\ref{avg_power}). Since RRH operates in the sleep mode with less power consumption than in the active mode, the activation algorithm has a higher power reduction than the sleep algorithm \cite{Amer2021}. At the same power reduction as the activation algorithm, our proposed algorithm further reduces the downlink power by utilizing the power offset in Equation (\ref{RRH_oft_power}) based on the UE throughput requirement. Hence, our proposed algorithm can enjoy much higher power reduction. Similar to the average power reduction in Equation (\ref{avg_power}), the average power offset for the C-RAN system is denoted as
\begin{equation}\label{avg_power_offset}
    \begin{gathered}
      \Delta {\overline P _{{\rm{Offset}}}} = 10 \cdot {\log _{10}}\left( {\frac{{\sum\limits_{b = 0}^{B - 1} {{{10}^{\frac{{P_{b,u}^\delta }}{{10}}}}} }}{B}} \right).
    \end{gathered}
\end{equation} 
For our proposed algorithm, \textbf{Fig. }\ref{3Power_reduction_UE}-(2) shows the downward trends in the average power offset, average power reduction, and average interference reduction when the number of activated RRHs increases, which explains the downward trend of the average power reduction of our proposed algorithm in \textbf{Fig. }\ref{3Power_reduction_UE}-(1). Moreover, the average interference reduction for the C-RAN system is denoted as
\begin{equation}\label{avg_power_ITF}
    \begin{gathered}
      \begin{array}{l}\!\!\!\!\!\!
        \Delta {\overline P _{{\rm{Interference \  reduction}}}} = \\        \qquad 
        10\!\cdot\!{\log _{10}}\!\!\left(\!\!{\frac{{\sum\limits_{b^{'} = 0, b^{'}\ne b }^{B - 1} {\sum\limits_{{u^{'}} = 0,{u^{'}} \ne u}^{U - 1} {\left( {{{10}^{\frac{{{P_{\max }}}}{{10}}}} - {{10}^{\frac{{{P_{b^{'},{u^{'}}}}}}{{10}}}}} \right)} } }}{B}}\!\!\right)
      \end{array}\!\!\!\!.
    \end{gathered}
\end{equation}
The average interference for the C-RAN system is denoted as
\begin{equation}\label{avg_power_ITF}
    \begin{gathered}
      \begin{array}{l}
        {\overline P _{{\rm{Interference}}}} = \\        \qquad \quad 
        10 \cdot {\log _{10}}\left( {\frac{{\sum\limits_{b^{'} = 0, b^{'}\ne b }^{B - 1} {\sum\limits_{{u^{'}} = 0,{u^{'}} \ne u}^{U - 1} {\left( {{{10}^{\frac{{{P_{b^{'},{u^{'}}}}}}{{10}}}}} \right)} } }}{B}} \right)
      \end{array}.
    \end{gathered}
\end{equation}
It is intuitive that the increasing number of activated RRHs can result in increasing the system interference, which definitely shrinks the power offset for each RRH in C-RAN. The decreased power offset leads to decreasing power reduction, resulting in decreasing interference reduction. 
\par
In \textbf{Fig. }\ref{4Power_reduction_weight}, the average power reduction for the C-RAN system is depicted based on diverse values of $(w_0, w_1)$. \textbf{Fig. }\ref{4Power_reduction_weight}-(1) shows that the average power reduction of our proposed algorithm, activation algorithm, and sleep algorithm almost remain unchanged when the value of weight $(w_0, w_1)$ varies. Our proposed algorithm enjoys a superior average power reduction compared to the activation and sleep schemes. The trends of average power offset, average power reduction, and average interference in \textbf{Fig. }\ref{4Power_reduction_weight}-(2) also almost remain unchanged when the value of weight $(w_0, w_1)$ varies. Power offset is the key determinant of power reduction based on Equation (15). The variation trends of power reduction and power offset are almost synchronous. Meanwhile, the operators $\left[ {0, \ldots ,\left\lfloor {P_{b,u}^\delta } \right\rfloor } \right]$ and $\left[ {0, \ldots ,P_\Upsilon ^\delta } \right]$ in Equation (15) can cause power reduction to be slightly smaller than power offset. Hence, the two parameters denoted by$(w_0, w_1)$ have a slight effect on our proposed algorithms. 
\begin{figure*}[htbp]
  \centering{\includegraphics[width=7.15 in]{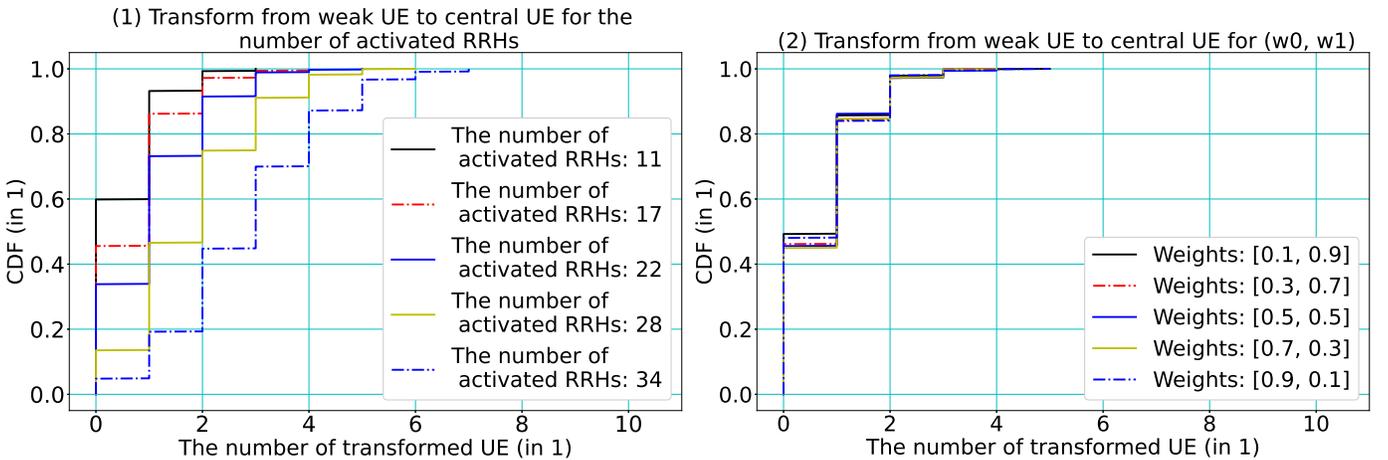}}
  \caption{Variation of UE satisfaction transition on different number of activated RRHs and diverse values of $(w_0, w_1)$.}\label{6Power_satisfy}
\end{figure*}
\par
\textbf{Fig. }\ref{5RSRPOft_action_ITF_analysis} depicts the power offset, power reduction, interference reduction, and throughput loss when $(w_0, w_1)=(0.5, 0.5)$ and the number of activated RRHs is 11. The values of average power offset, power reduction, interference reduction, and  throughput loss are, respectively, 23.27dB, 13.01dB, 8.66dB, and 6.60Mb/s. The average interference reduction is almost half of power reduction, indicating a network-level throughput loss is hard to be avoided when optimizing power reduction. The average power reduction is almost half of the average power offset, indicating that there is still great potential for power reduction to be explored. When the number of activated RRHs increases, the power offset, power reduction, interference reduction, and throughput loss are specifically analyzed in Section \ref{Transform_central_weak_UE}. 
\par
From \textbf{Fig. }\ref{3Power_reduction_UE} and \textbf{Fig. }\ref{4Power_reduction_weight}, we can learn that the number of activated RRHs has an important effect on the power offset, power reduction, and interference reduction, and that the variation of $(w_0, w_1)$ has a slight effect on these reductions. We will analyze the variations with increasing number of activated RRHs in Section \ref{Transform_central_weak_UE} . 
\subsection{Increasing Number of Activated RRHs}\label{Transform_central_weak_UE}
In C-RAN, the activated RRHs serving central UEs perform power reduction, while weak UEs are free from power reduction. Weak UEs will enjoy throughput increment and can possibly be transformed from unsatisfied state to satisfied state based on the comparison with their own throughput requirements. \textbf{Fig. }\ref{6Power_satisfy} shows CDFs of satisfaction variation of UEs when the number of activated RRHs and value of $(w_0, w_1)$ vary. It can be learned that the increasing number of activated RRHs contributes to the transformation from weak UEs to central UEs and the variation of $(w_0, w_1)$ has a slight effect on the transformation between weak UEs and central UEs. Hence, we mainly analyze the impact of the increasing number of activated RRHs by plotting the CDF.
\begin{figure*}[htbp]
  \centering{\includegraphics[width=7.15 in]{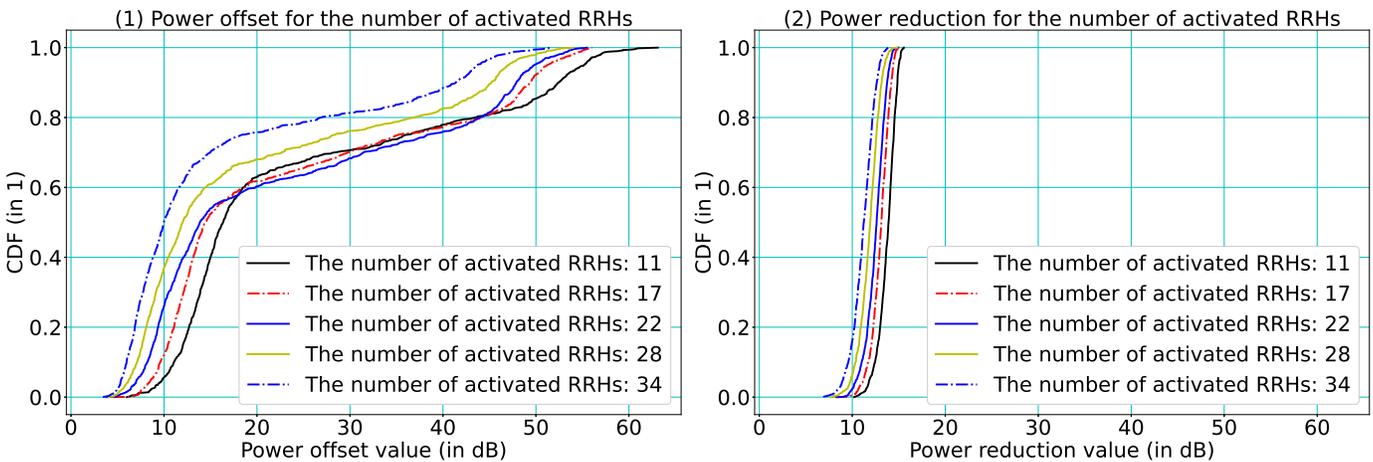}}
  \caption{Variation of power offset and reduction on different number of activated RRHs.}\label{2Power_Power_offset_reduction_dec}
\end{figure*}
\begin{figure*}[htbp]
  \centering{\includegraphics[width=7.15 in]{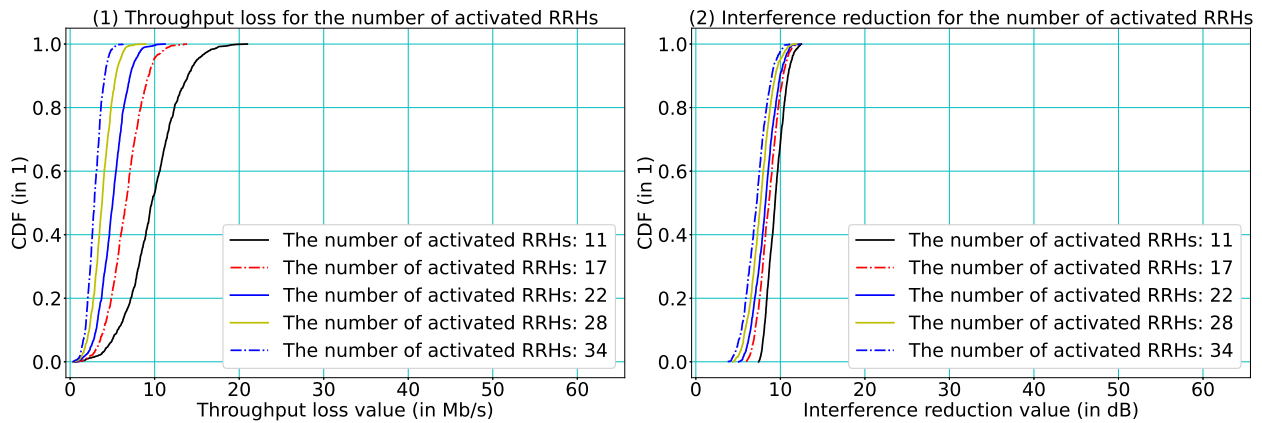}}
  \caption{Variation of throughput loss and interference reduction on different number of activated RRHs.}\label{8Power_THR_ITF_dec}
\end{figure*}
\par
When increasing the number of activated RRHs, the CDFs of power offset, power reduction, interference reduction, and throughput loss are depicted in four subgraphs in \textbf{Fig. }\ref{2Power_Power_offset_reduction_dec} and \textbf{Fig. }\ref{8Power_THR_ITF_dec}. For the convenience of comparison, we have unified the x-axis coordinates of these four subgraphs in \textbf{Fig. }\ref{2Power_Power_offset_reduction_dec} and \textbf{Fig. }\ref{8Power_THR_ITF_dec}. In any subgraph, the CDF curve corresponding to the larger number of activated RRHs is closer to the side with the smaller x-axis value in the graph. This implies that the increasing number of activated RRHs shrinks the power offset, power reduction, interference reduction, and throughput loss. The comparison between \textbf{Fig. }\ref{2Power_Power_offset_reduction_dec}-(1) and \textbf{Fig. }\ref{2Power_Power_offset_reduction_dec}-(2) shows that the CDF curves corresponding to the power reduction are closer to the side with the smaller x-axis value than the CDF curves of power offset for different numbers of activated RRHs. The comparison between \textbf{Fig. }\ref{8Power_THR_ITF_dec}-(2) and \textbf{Fig. }\ref{2Power_Power_offset_reduction_dec}-(2) shows that the CDF curves corresponding to the interference reduction are closer to the side with the smaller x-axis value than the CDF curves of power reduction for different numbers of activated RRHs. All the CDF curves trends are completely consistent with \textbf{Fig. }\ref{3Power_reduction_UE}-(2). 
\begin{table*}[t]
\caption{The average statistics}
\begin{center}
\begin{tabular}{|c|c|c|c|c|c|c|c|c|c|c|c|c|c|c|}
\hline
\multicolumn{7}{|c|}{\textbf{Scenario I for $\left( {\left| {\mathcal B} \right|,{D_{b,b^{'}}}} \right) = \left( {57,200} \right)$}} &\!\!\!\!\!\!\!& \multicolumn{6}{c|}{\textbf{Scenario I for $\left( {\left| {\mathcal B} \right|,{D_{b,b^{'}}}} \right) = \left( {57,200} \right)$}} \\
\hline
\multicolumn{2}{|c|}{\textbf{\!\!\!\!\!The number of activated RRHs\!\!\!\!\!}} & \textbf{11} & \textbf{17} & \textbf{22} & \textbf{28}& \textbf{34} &\!\!\!\!\!\!\!& \textbf{\!\!\!\!Values for $(w_0,w_1)$\!\!\!\!} & \textbf{\!\!\!\!(0.1, 0.9)\!\!\!\!} & \textbf{\!\!\!\!(0.3, 0.7)\!\!\!\!} & \textbf{\!\!\!\!(0.5, 0.5)\!\!\!\!} & \textbf{\!\!\!\!(0.7, 0.3)\!\!\!\!} & \textbf{\!\!\!\!(0.9, 0.1)\!\!\!\!} \\
\hline
\multicolumn{2}{|c|}{\textbf{Power offset}} &\!\!24.6\!\!&\!\!23.3\!\!&\!\!22.8\!\!&\!\!19.3\!\!&\!\!16.1 &\!\!\!\!\!\!\!&\textbf{Power offset}\!\!&\!\!22.3\!\!&\!\!23.5\!\!&\!\!23.3\!\!&\!\!23.6\!\!&\!\!23.4\!\!\\
\hline
\multicolumn{2}{|c|}{\textbf{Power reduction}} &\!\!13.7\!\!&\!\!13.0\!\!&\!\!12.5\!\!&\!\!11.8\!\!&\!\!11.2\!\!&\!\!\!\!\!\!\!&{\textbf{Power reduction}} &\!\!13.0\!\!&\!\!13.0\!\!&\!\!13.0\!\!&\!\!13.0\!\!&\!\!13.0\!\!\\
\hline
\multicolumn{2}{|c|}{\textbf{\!\!\!Interference reduction\!\!\!}} &\!\!9.4\!\!&\!\!8.7\!\!&\!\!8.2\!\!&\!\!7.7\!\!&\!\!7.2\!\!&\!\!\!\!\!\!\!& \!\!{\textbf{Interference reduction}} &\!\!8.7\!\!&\!\!8.7\!\!&\!\!8.7\!\!&\!\!8.7\!\!&\!\!8.7\!\!\\
\hline
\multicolumn{2}{|c|}{\textbf{Throughout loss}} &\!\!9.8\!\!&\!\!6.6\!\!&\!\!5.1\!\!&\!\!3.8\!\!&\!\!2.9\!\! &\!\!\!\!\!\!\!& {\textbf{Throughout loss}} &\!\!6.5\!\!&\!\!6.5\!\!&\!\!6.6\!\!&\!\!6.5\!\!&\!\!6.6\!\!\\
\hline
\multicolumn{2}{|c|}{\textbf{Weak UEs $\to$ central UEs}} & 0.5 & 0.7 & 1.0 & 1.8 & 2.8  &\!\!\!\!\!\!\!& {\textbf{Weak UEs $\to$ central UEs}} &\!\!0.7\!\!&\!\!0.7\!\!&\!\!0.7\!\!&\!\!0.7\!\!&\!\!0.7\!\!\\
\hline
\multicolumn{2}{|c|}{\textbf{Central UEs $\to$ weak UEs}} & 0 & 0 & 0 & 0 & 0 &\!\!\!\!\!\!\!& {\textbf{Central UEs $\to$ weak UEs}} & 0 & 0 & 0 & 0 & 0   \\
\hline
\multicolumn{2}{|c|}{\textbf{Iterations per epoch}} &\!\!14.1\!\!&\!\!16.3\!\!&\!\!18.2\!\!&\!\!20.2\!\!&\!\!22.0\!\! &\!\!\!\!\!\!\!& {\textbf{Iterations per epoch}} &\!\!16.2\!\!&\!\!16.0\!\!&\!\!16.3\!\!&\!\!16.1\!\!&\!\!15.6\!\! \\
\hline
\multicolumn{7}{|c|}{\textbf{Scenario II for $\left( {\left| {\mathcal B} \right|,{D_{b,b^{'}}}} \right) = \left( {57,300} \right)$}} &\!\!\!\!\!\!\!& \multicolumn{6}{c|}{\textbf{Scenario III for $\left( {\left| {\mathcal B} \right|,{D_{b,b^{'}}}} \right) = \left( {111,200} \right)$}} \\
\hline
\multicolumn{2}{|c|}{\textbf{\!\!\!The number of activated RRHs\!\!\!}} & \textbf{11} & \textbf{17} & \textbf{22} & \textbf{28}& \textbf{34} &\!\!\!\!\!\!\!& \textbf{\!\!\!\!The number of activated RRHs\!\!\!\!} & \textbf{22} & \textbf{33} & \textbf{44} & \textbf{55} & \textbf{66} \\
\hline
\multicolumn{2}{|c|}{\textbf{Power offset}} &\!\!18.8\!\!&\!\!18.0\!\!&\!\!16.1\!\!&\!\!14.1\!\!&\!\!11.1 &\!\!\!\!\!\!\!&\textbf{Power offset}\!\!&\!\!31.5\!\!&\!\!29.9\!\!&\!\!27.7\!\!&\!\!24.6\!\!&\!\!20.8\!\!\\
\hline
\multicolumn{2}{|c|}{\textbf{Power reduction}} &\!\!12.8\!\!&\!\!12.0\!\!&\!\!11.1\!\!&\!\!10.5\!\!&\!\!9.7\!\!&\!\!\!\!\!\!\!&{\textbf{Power reduction}} &\!\!13.9\!\!&\!\!13.3\!\!&\!\!12.7\!\!&\!\!12.1\!\!&\!\!11.5\!\!\\
\hline
\multicolumn{2}{|c|}{\textbf{\!\!\!Interference reduction\!\!\!}} &\!\!8.5\!\!&\!\!7.5\!\!&\!\!6.9\!\!&\!\!6.4\!\!&\!\!5.8\!\!&\!\!\!\!\!\!\!& \!\!{\textbf{Interference reduction}} &\!\!9.7\!\!&\!\!9.1\!\!&\!\!8.6\!\!&\!\!8.1\!\!&\!\!7.6\!\!\\
\hline
\multicolumn{2}{|c|}{\textbf{Throughout loss}} &\!\!8.8\!\!&\!\!6.2\!\!&\!\!4.6\!\!&\!\!3.4\!\!&\!\!2.6\!\! &\!\!\!\!\!\!\!& {\textbf{Throughout loss}} &\!\!9.3\!\!&\!\!6.5\!\!&\!\!4.7\!\!&\!\!3.7\!\!&\!\!2.8\!\!\\
\hline
\multicolumn{2}{|c|}{\textbf{Weak UEs $\to$ central UEs}} & 0.2 & 0.4 & 0.6 & 1.3 & 2.3  &\!\!\!\!\!\!\!& {\textbf{Weak UEs $\to$ central UEs}} &\!\!1.2\!\!&\!\!1.6\!\!&\!\!2.3\!\!&\!\!3.4\!\!&\!\!5.1\!\!\\
\hline
\multicolumn{2}{|c|}{\textbf{Central UEs $\to$ weak UEs}} & 0 & 0 & 0 & 0 & 0 &\!\!\!\!\!\!\!& {\textbf{Central UEs $\to$ weak UEs}} & 0 & 0 & 0 & 0 & 0   \\
\hline
\multicolumn{2}{|c|}{\textbf{Iterations per epoch}} &\!\!13.9\!\!&\!\!16.6\!\!&\!\!18.5\!\!&\!\!20.7\!\!&\!\!22.4\!\! &\!\!\!\!\!\!\!& {\textbf{Iterations per epoch}} &\!\!15.9\!\!&\!\!18.3\!\!&\!\!20.9\!\!&\!\!22.9\!\!&\!\!24.8\!\! \\
\hline
\end{tabular}
\label{avg_sta}
\end{center}
\end{table*}
The activation and sleep schemes are two different power control mechanisms that make RRHs into two different dormant states. The activation and sleep schemes have differences in power savings, while they have the same network throughput. \textbf{Fig. }\ref{8Power_THR_ITF_dec}-(1) shows that our proposed algorithm brings different network throughput loss for different numbers of activated RRHs compared to the activation and sleep schemes. The increasing number of activated RRHs is useful to reduce throughput loss due to better interference coordination. Although our proposed algorithm can cause network throughput loss, it improves satisfaction of UEs as illustrated in \textbf{TABLE} \ref{avg_sta}. 
Moreover, the average power offset, power reduction, interference reduction, and throughput loss are given in \textbf{TABLE} \ref{avg_sta} when the number of activated RRHs and $(w_0, w_1)$ vary. The statistical data in \textbf{TABLE} \ref{avg_sta} also indicates that the number of activated RRHs has an important effect on the variations of power offset, power reduction, interference reduction, and throughput loss, and that the value of $(w_0, w_1)$ has a slight effect on these variations. \textbf{TABLE} \ref{avg_sta} also provides the number of iterations related to the computational complexity of our proposed algorithm, which will be discussed in Section \ref{discussion_optimization_complexity}. 
To justify the scalability and portability of our proposed algorithm, we show the network data statistics under three typical scenario setups in \textbf{TABLE} \ref{avg_sta}. Based on the radio communication model in Section II-A, the key network factor of the scenario setup is defined by the pair $\left( {\left| {\mathcal B} \right|,{D_{b,b^{'}}}} \right)$, where $\left| {\mathcal B} \right|$ is the network scale and $D_{b,b^{'}}$ is the distance between RRHs. As the size of network scales up, the network interference increases. This not only leads to the higher computational complexity, but also increases the optimization space for the power offset and power reduction. Conversely, increasing the distance between RRHs decreases the network interference, and reduces the optimization space for the power offset and power reduction. However, \textbf{TABLE} \ref{avg_sta} shows similar network data statistics for different scenario setups, implying that our proposed algorithm can adapt to different downlink C-RAN scenario setups.
\subsection{Discussion on Optimization Complexity}\label{discussion_optimization_complexity}
\begin{figure*}[t]
  \centering{\includegraphics[width=7.15 in]{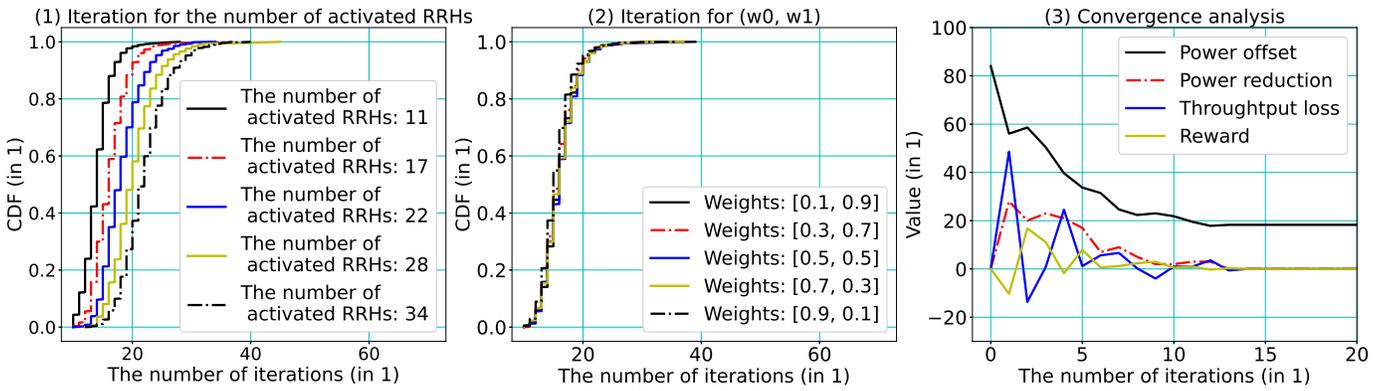}}
  \caption{Iteration on different numbers of activated RRHs and diverse values of $(w_0, w_1)$, and convergence analysis.}\label{9iteration_analysis}
\end{figure*}

To quantify the complexity of our proposed algorithm, \textbf{Fig. }\ref{9iteration_analysis} shows the iteration and convergence analysis for different numbers of activated RRHs and diverse values of $(w_0, w_1)$. \textbf{Fig. }\ref{9iteration_analysis}-(1) shows the CDF curves when the number of activated RRHs increases. 
We can learn that the increasing number of activated RRHs can result in more iterations, because it expands the scale of C-RAN optimization. However, the variation of $(w_0, w_1)$ does not impact the scale of network optimization. Hence, the value of $(w_0, w_1)$ has a slight effect on the number of iterations, which is illustrated in \textbf{Fig. }\ref{9iteration_analysis}-(2). When $(w_0, w_1)=(0.5, 0.5)$ and the number of activated RRHs is 11, \textbf{Fig. }\ref{9iteration_analysis}-(3) shows an example of the convergence process for the power offset, power reduction, interference reduction, and reward defined in Equation (\ref{immediate_reward}), indicating that these four metrics converge synchronously and quickly. Meanwhile, the variation trends of the power offset and power reduction are completely synchronized. According to the statistical iterations in \textbf{TABLE} \ref{avg_sta} and \textbf{Fig. }\ref{9iteration_analysis}, our proposed algorithm enjoys a low computational complexity.
\par
Based on the above numerical results, our proposed algorithm not only achieves better power saving in comparison with the activation and sleep schemes, but also enjoys a low computational complexity. Hence, our proposed algorithm can transform some weak UEs to central UEs while satisfying the throughput requirement of central UEs. It can be learned that the increasing number of activated RRHs results in a modicum increase in the number of iterations, and that the variation of $(w_0, w_1)$ has a slight effect on the number of iterations.
\section{CONCLUSION}\label{Conclu}
In this paper, we proposed a novel static deep Q-learning algorithm to maximize the long-term downlink throughput and power saving in C-RAN. Based on the unpredictable differentiated traffic demands of UEs, the BBU simultaneously optimizes the downlink power of activated RRHs without degrading UEs' throughput requirements. To achieve precise downlink power reduction, we defined the power offset for RRHs based on the UE throughput requirement. We formulated a dual objective optimization problem with UEs' throughput requirements. Integrating deep learning and Q-learning, we designed a novel static deep Q-learning to maximize the long-term downlink throughput and power saving in C-RAN. In our proposed algorithm, each UE corresponds to one Q-table. RSRP at the UE and power reduction at the RRH are, respectively, defined as the state and action for each Q-table. The additive value of power reduction and throughput loss is defined as the reward. To maximize the accumulative reward, our proposed algorithm performs power reduction through continuous environmental interactions. Simulation results show that our proposed algorithm not only enjoys a better average power reduction than the activation and sleep schemes, but also enjoys a low computational complexity. 

\bibliographystyle{IEEEtran}
\bibliography{bibtex}

\begin{thebibliography}{10}
\providecommand{\url}[1]{#1}
\csname url@samestyle\endcsname
\providecommand{\newblock}{\relax}
\providecommand{\bibinfo}[2]{#2}
\providecommand{\BIBentrySTDinterwordspacing}{\spaceskip=0pt\relax}
\providecommand{\BIBentryALTinterwordstretchfactor}{4}
\providecommand{\BIBentryALTinterwordspacing}{\spaceskip=\fontdimen2\font plus
\BIBentryALTinterwordstretchfactor\fontdimen3\font minus
  \fontdimen4\font\relax}
\providecommand{\BIBforeignlanguage}[2]{{%
\expandafter\ifx\csname l@#1\endcsname\relax
\typeout{** WARNING: IEEEtran.bst: No hyphenation pattern has been}%
\typeout{** loaded for the language `#1'. Using the pattern for}%
\typeout{** the default language instead.}%
\else
\language=\csname l@#1\endcsname
\fi
#2}}
\providecommand{\BIBdecl}{\relax}
\BIBdecl

\bibitem{Gvozdenovic2023IoT}
S.~Gvozdenovic, J.~K. Becker, J.~Mikulskis, and D.~Starobinski, ``{IoT-Scan:
  Network Reconnaissance for the Internet of Things},'' \emph{IEEE Internet of
  Things Journal}, pp. 1--1, 2023.

\bibitem{Chen2023JSAC}
W.~Chen, X.~Lin, J.~Lee, A.~Toskala, S.~Sun, C.~F. Chiasserini, and L.~Liu,
  ``{5G-Advanced Toward 6G: Past, Present, and Future},'' \emph{IEEE Journal on
  Selected Areas in Communications}, vol.~41, no.~6, pp. 1592--1619, 2023.

\bibitem{Chang2017Sensors}
Y.~Chang, H.~Tang, Y.~Cheng, Q.~Zhao, X.~Yuan, and B.~Li, ``{Dynamic
  Hierarchical Energy-Efficient Method Based on Combinatorial Optimization for
  Wireless Sensor Networks},'' \emph{Sensors}, vol.~17, no.~7, 2017.

\bibitem{SaxenaN2016JSAC}
N.~Saxena, A.~Roy, and H.~Kim, ``{Traffic-Aware Cloud RAN: A Key for Green 5G
  Networks},'' \emph{IEEE J. Sel. Areas Commun.}, vol.~34, no.~4, pp.
  1010--1021, 2016.

\bibitem{pompili2016elastic}
D.~Pompili, A.~Hajisami, and T.~X. Tran, ``{Elastic resource utilization
  framework for high capacity and energy efficiency in cloud RAN},'' \emph{IEEE
  Commun. Mag.}, vol.~54, no.~1, pp. 26--32, 2016.

\bibitem{peng2016recent}
M.~Peng, Y.~Sun, X.~Li, Z.~Mao, and C.~Wang, ``{Recent advances in cloud radio
  access networks: System architectures, key techniques, and open issues},''
  \emph{IEEE Commun. Surveys Tuts.}, vol.~18, no.~3, pp. 2282--2308, 2016.

\bibitem{wu2012green}
J.~Wu, ``{Green wireless communications: from concept to reality},'' \emph{IEEE
  Wireless Commun.}, vol.~19, no.~4, pp. 4--5, 2012.

\bibitem{marotta2017characterizing}
M.~A. Marotta, H.~Ahmadi, J.~Rochol, L.~DaSilva, and C.~B. Both,
  ``{Characterizing the relation between processing power and distance between
  BBU and RRH in a cloud RAN},'' \emph{” IEEE Wireless Commun. Lett.},
  vol.~7, no.~3, pp. 472--475, 2017.

\bibitem{ChangY2018LCOMM}
Y.~Chang, X.~Yuan, B.~Li, D.~Niyato, and N.~Al-Dhahir, ``{A Joint Unsupervised
  Learning and Genetic Algorithm Approach for Topology Control in
  Energy-Efficient Ultra-Dense Wireless Sensor Networks},'' \emph{IEEE Commun.
  Lett.}, vol.~22, no.~11, pp. 2370--2373, 2018.

\bibitem{ICC2021Yue}
X.~Yue, K.~Sun, W.~Huang, X.~Liu, and H.~Zhang, ``{Beamforming Design and BBU
  Computation Resource Allocation for Power Minimization in Green C-RAN},'' in
  \emph{Proc. IEEE Int. Conf. Commun. (ICC)}, 2021, pp. 1--6.

\bibitem{ICT2021The}
C.~Freitag, M.~Berners-Lee, K.~Widdicks, B.~Knowles, G.~Blair, and A.~Friday,
  ``{The climate impact of ICT: A review of estimates, trends and
  regulations},'' 2021.

\bibitem{mao2021ai}
B.~Mao, F.~Tang, Y.~Kawamoto, and N.~Kato, ``{AI models for green
  communications towards 6G},'' \emph{IEEE Commun. Surveys Tuts.}, vol.~24,
  no.~1, pp. 210--247, 2021.

\bibitem{TACT2014_RLi}
R.~{Li}, Z.~{Zhao}, X.~{Chen}, J.~{Palicot}, and H.~{Zhang}, ``{TACT: A
  Transfer Actor-Critic Learning Framework for Energy Saving in Cellular Radio
  Access Networks},'' \emph{IEEE Trans. Wireless Commun.}, vol.~13, no.~4, pp.
  2000--2011, 2014.

\bibitem{mahapatra2015energy}
R.~Mahapatra, Y.~Nijsure, G.~Kaddoum, N.~U. Hassan, and C.~Yuen, ``{Energy
  efficiency tradeoff mechanism towards wireless green communication: A
  survey},'' \emph{IEEE Commun. Surveys Tuts.}, vol.~18, no.~1, pp. 686--705,
  2015.

\bibitem{zhang2016fundamental}
S.~Zhang, Q.~Wu, S.~Xu, and G.~Y. Li, ``{Fundamental Green Tradeoffs:
  Progresses, Challenges, and Impacts on 5G Networks},'' \emph{IEEE Commun.
  Surveys Tuts.}, vol.~19, no.~1, pp. 33--56, 2017.

\bibitem{BBDai2016JSAC}
B.~Dai and W.~Yu, ``{Energy Efficiency of Downlink Transmission Strategies for
  Cloud Radio Access Networks},'' \emph{IEEE J. Sel. Areas Commun.}, vol.~34,
  no.~4, pp. 1037--1050, 2016.

\bibitem{sigwele2020energy}
T.~Sigwele, Y.~F. Hu, and M.~Susanto, ``{Energy-efficient 5G cloud RAN with
  virtual BBU server consolidation and base station sleeping},'' \emph{J. Netw.
  Comput.}, vol. 177, p. 107302, 2020.

\bibitem{TCOMM.2021.3091133}
Z.~Yang, M.~Chen, W.~Saad, and M.~Shikh-Bahaei, ``{Optimization of Rate
  Allocation and Power Control for Rate Splitting Multiple Access (RSMA)},''
  \emph{IEEE Trans. Commun.}, vol.~69, no.~9, pp. 5988--6002, 2021.

\bibitem{Mao2022COMST319}
Y.~Mao, O.~Dizdar, B.~Clerckx, R.~Schober, P.~Popovski, and H.~V. Poor,
  ``{Rate-Splitting Multiple Access: Fundamentals, Survey, and Future Research
  Trends},'' \emph{IEEE Commun. Surveys Tuts.}, vol.~24, no.~4, pp. 2073--2126,
  2022.

\bibitem{TCM2021Teng}
W.~Teng, M.~Sheng, X.~Chu, K.~Guo, J.~Wen, and Z.~Qiu, ``{Joint Optimization of
  Base Station Activation and User Association in Ultra Dense Networks Under
  Traffic Uncertainty},'' \emph{IEEE Trans. Commun.}, vol.~69, no.~9, pp.
  6079--6092, 2021.

\bibitem{RTao2019TWC}
R.~Tao, W.~Liu, X.~Chu, and J.~Zhang, ``{An Energy Saving Small Cell Sleeping
  Mechanism With Cell Range Expansion in Heterogeneous Networks},'' \emph{IEEE
  Trans. Wireless Commun.}, vol.~18, no.~5, pp. 2451--2463, 2019.

\bibitem{qin2020green}
M.~Qin, W.~Wu, Q.~Yang, R.~Zhang, N.~Cheng, H.~Zhou, R.~R. Rao, and X.~Shen,
  ``{Green-oriented dynamic resource-on-demand strategy for multi-RAT wireless
  networks powered by heterogeneous energy sources},'' \emph{IEEE Trans.
  Wireless Commun.}, vol.~19, no.~8, pp. 5547--5560, 2020.

\bibitem{TVT.2017.2719404}
M.~Oikonomakou, A.~Antonopoulos, L.~Alonso, and C.~Verikoukis, ``{Evaluating
  Cost Allocation Imposed by Cooperative Switching Off in Multioperator Shared
  HetNets},'' \emph{IEEE Trans. Veh. Technol.}, vol.~66, no.~12, pp.
  11\,352--11\,365, 2017.

\bibitem{TVT2016Bousia}
A.~Bousia, E.~Kartsakli, A.~Antonopoulos, L.~Alonso, and C.~Verikoukis,
  ``{Multiobjective Auction-Based Switching-Off Scheme in Heterogeneous
  Networks: To Bid or Not to Bid?}'' \emph{IEEE Trans. Veh. Technol.}, vol.~65,
  no.~11, pp. 9168--9180, 2016.

\bibitem{chang2017distributed}
Y.~Chang, H.~Tang, B.~Li, and X.~Yuan, ``Distributed joint optimization routing
  algorithm based on the analytic hierarchy process for wireless sensor
  networks,'' \emph{IEEE Communications Letters}, vol.~21, no.~12, pp.
  2718--2721, 2017.

\bibitem{letaief2019roadmap}
K.~B. Letaief, W.~Chen, Y.~Shi, J.~Zhang, and Y.-J.~A. Zhang, ``{The roadmap to
  6G: AI empowered wireless networks},'' \emph{IEEE Commun. Mag.}, vol.~57,
  no.~8, pp. 84--90, 2019.

\bibitem{YChangACCESS2019}
Y.~Chang, X.~Yuan, B.~Li, D.~Niyato, and N.~Al-Dhahir,
  ``{Machine-Learning-Based Parallel Genetic Algorithms for Multi-Objective
  Optimization in Ultra-Reliable Low-Latency WSNs},'' \emph{IEEE Access},
  vol.~7, pp. 4913--4926, 2019.

\bibitem{gu2021knowledge}
Z.~Gu, C.~She, W.~Hardjawana, S.~Lumb, D.~McKechnie, T.~Essery, and B.~Vucetic,
  ``{Knowledge-assisted deep reinforcement learning in 5G scheduler design:
  From theoretical framework to implementation},'' \emph{IEEE J. Sel. Areas
  Commun.}, vol.~39, no.~7, pp. 2014--2028, 2021.

\bibitem{xu2020buffer}
C.~Xu, J.~Wang, T.~Yu, C.~Kong, Y.~Huangfu, R.~Li, Y.~Ge, and J.~Wang,
  ``{Buffer-aware wireless scheduling based on deep reinforcement learning},''
  in \emph{Proc. IEEE Wireless Commun. Netw. Conf. (WCNC)}.\hskip 1em plus
  0.5em minus 0.4em\relax IEEE, 2020, pp. 1--6.

\bibitem{wang2019deep}
J.~Wang, C.~Xu, Y.~Huangfu, R.~Li, Y.~Ge, and J.~Wang, ``{Deep reinforcement
  learning for scheduling in cellular networks},'' in \emph{Proc. 11th Int.
  Conf. Wireless Commun. Signal Process. (WCSP)}.\hskip 1em plus 0.5em minus
  0.4em\relax IEEE, 2019, pp. 1--6.

\bibitem{Book_RL}
R.~S. {Sutton} and A.~G. {Barto}, \emph{{Reinforcement Learning: An
  Introduction}}.\hskip 1em plus 0.5em minus 0.4em\relax Massachusetts London,
  England: The MIT Press Cambridge, 2018.

\bibitem{zhang2022q}
L.~Zhang, X.~Ma, Z.~Zhuang, H.~Xu, V.~Sharma, and Z.~Han, ``{Q-Learning Aided
  Intelligent Routing With Maximum Utility in Cognitive UAV Swarm for Emergency
  Communications},'' \emph{IEEE Trans. Veh. Technol.}, vol.~72, no.~3, pp.
  3707--3723, 2022.

\bibitem{ICCCS2021Ding}
Y.~Ding, Y.~Zhao, Y.~Gao, and R.~Zhang, ``{Q-Learning Quantum Ant Colony
  Routing Algorithm for Micro-Nano Satellite Network},'' in \emph{in Proc. IEEE
  6th Int. Conf. Comput. Commun. Syst. (ICCCS)}, 2021, pp. 949--954.

\bibitem{JCC.2021.08.016}
Y.~Chen, K.~Zheng, X.~Fang, L.~Wan, and X.~Xu, ``{QMCR: A Q-learning-based
  multi-hop cooperative routing protocol for underwater acoustic sensor
  networks},'' \emph{China Commun.}, vol.~18, no.~8, pp. 224--236, 2021.

\bibitem{JIOT.2021.3098331}
S.~Zou, W.~Wang, W.~Ni, L.~Wang, and Y.~Tang, ``{Efficient Orchestration of
  Virtualization Resource in RAN Based on Chemical Reaction Optimization and
  Q-Learning},'' \emph{IEEE Internet Things J.}, vol.~9, no.~5, pp. 3383--3396,
  2022.

\bibitem{TNSM.2023.3245544}
A.~Ndikumana, K.~K. Nguyen, and M.~Cheriet, ``{Federated Learning Assisted Deep
  Q-Learning for Joint Task Offloading and Fronthaul Segment Routing in Open
  RAN},'' \emph{IEEE Trans. Netw. Service Manage.}, pp. 1--1, 2023.

\bibitem{OJCOMS.2022.3219014}
A.~A. hammadi, L.~Bariah, S.~Muhaidat, M.~Al-Qutayri, P.~C. Sofotasios, and
  M.~Debbah, ``{Deep Q-Learning-Based Resource Allocation in NOMA Visible Light
  Communications},'' \emph{IEEE Open J. Commun. Soc.}, vol.~3, pp. 2284--2297,
  2022.

\bibitem{lecun2015deep}
Y.~LeCun, Y.~Bengio, and G.~Hinton, ``{Deep learning},'' \emph{nature}, vol.
  521, no. 7553, pp. 436--444, 2015.

\bibitem{TGCN2023Chang}
Y.~Chang, W.~Chen, J.~Li, J.~Liu, H.~Wei, Z.~Wang, and N.~Al-Dhahir,
  ``{Collaborative Multi-BS Power Management for Dense Radio Access Network
  Using Deep Reinforcement Learning},'' \emph{IEEE Trans. Green Commun. Netw.},
  vol.~7, no.~4, pp. 2104--2116, 2023.

\bibitem{WCLopez2019}
A.~V. Lopez, A.~Chervyakov, G.~Chance, S.~Verma, and Y.~Tang, ``{Opportunities
  and Challenges of mmWave NR},'' \emph{IEEE Wireless Commun.}, vol.~26, no.~2,
  pp. 4--6, 2019.

\bibitem{RoseHu2013Book}
R.~Q. Hu and Y.~Qian, \emph{{Connected-Mode Mobility in LTE Heterogeneous
  Networks}}, 2013, pp. 199--214.

\bibitem{ICC2020Marzouk}
F.~Marzouk, T.~Akhtar, I.~Politis, J.~P. Barraca, and A.~Radwan, ``{Power
  Minimizing BBU-RRH Group Based Mapping in C-RAN with Constrained Devices},''
  in \emph{Proc. IEEE Int. Conf. Commun. (ICC)}, 2020, pp. 1--7.

\bibitem{yifei2014application}
Y.~Yuan and L.~Zhu, ``{Application scenarios and enabling technologies of
  5G},'' \emph{China Commun.}, vol.~11, no.~11, pp. 69--79, 2014.

\bibitem{alzubaidi2022interference}
O.~T.~H. Alzubaidi, M.~N. Hindia, K.~Dimyati, K.~A. Noordin, A.~N.~A. Wahab,
  F.~Qamar, and R.~Hassan, ``{Interference challenges and management in B5G
  network design: A comprehensive review},'' \emph{Electronics}, vol.~11,
  no.~18, p. 2842, 2022.

\bibitem{Amer2021}
A.~A. Amer, I.~E. Talkhan, and T.~Ismail, ``{Optimal Power Consumption on
  Distributed Edge Services Under Non-Uniform Traffic with Dual Threshold
  Sleep/Active Control},'' in \emph{2021 3rd Novel Intelligent and Leading
  Emerging Sciences Conference (NILES)}, 2021, pp. 63--66.

\end{thebibliography}


\begin{IEEEbiography}[{\includegraphics[width=1.0in,height=1.20in,clip]{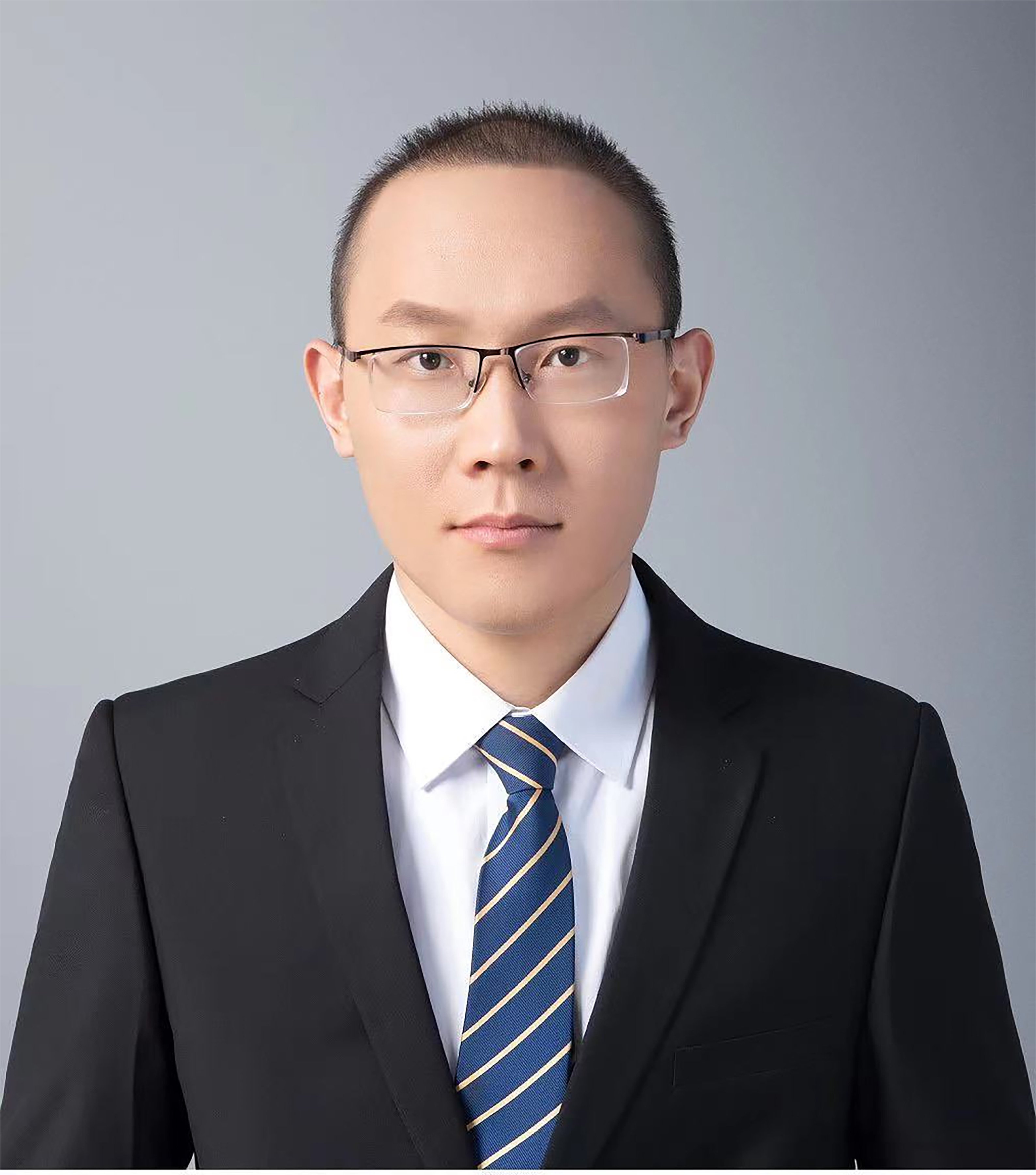}}]{Yuchao Chang} received the Ph.D. degree in Shanghai Institute of Microsystem and Information Technology from University of Chinese Academy of Sciences, China, in June 2019. From January 2018 to January 2019, he joined Prof. Naofal Al-Dhahir Group in University of Texas at Dallas as a Visiting Ph.D. Student, USA. 
His research interests include Artificial Intelligence and Machine Learning for Wireless, Green Internet of Things System, Green Communications, Green Computing, and Modeling and Algorithm Design. He is an Editor of International Journal of Machine Learning and Artificial Intelligence Systems and Applications. He serves as the Guest Editors for several journals, such as Electronics and Sustainability.
\end{IEEEbiography}

\begin{IEEEbiography}[{\includegraphics[width=1.0in,height=1.20in,clip]{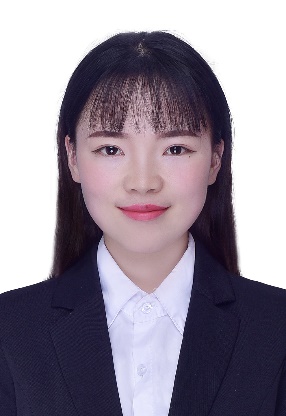}}]{Hongli Wang} received the Bachelor’s degree in telecommunications engineering from Jilin University, Changchun, China, in 2019. She is currently working toward the Master’s degree with Department of Electronics, Shanghai Jiao Tong University, Shanghai, China. Her research interests include federated learning and wireless AI.
\end{IEEEbiography}

\begin{IEEEbiography}[{\includegraphics[width=1.0in,height=1.00in,clip]{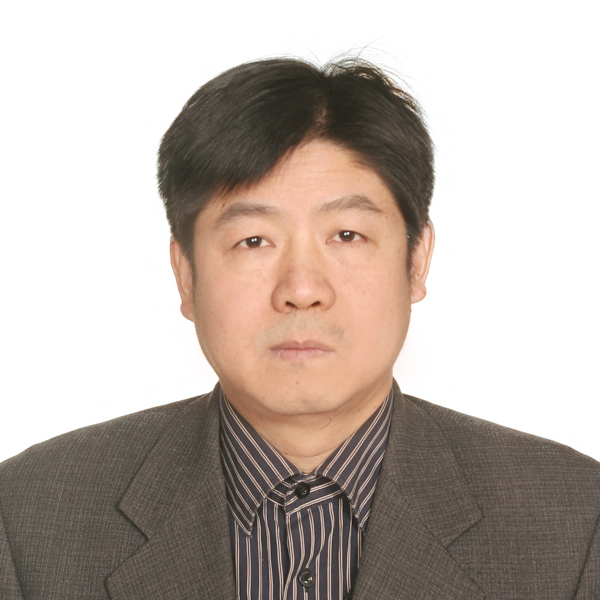}}] {Wen Chen} (Senior Member, IEEE) is currently a Tenured Professor with the Department of Electronic Engineering, Shanghai Jiao Tong University, China, where he is also the Director of the Broadband Access Network Laboratory. He has published more than 100 articles in IEEE journals and more than 120 papers in IEEE conferences, with citations more than 7000 in Google Scholar. His research interests include reconfigurable meta-surface, multiple access, wireless AI, and green networks. He is a fellow of the Chinese Institute of Electronics and a Distinguished Lecturer of the IEEE Communications Society and the IEEE Vehicular Technology Society. He is the Shanghai Chapter Chair of the IEEE Vehicular Technology Society. He is an Editor of IEEE Transactions on Wireless Communications, IEEE Transactions on Communications, IEEE Access, and IEEE Open Journal of Vehicular Technology.
\end{IEEEbiography}

\begin{IEEEbiography}[{\includegraphics[width=1.0in,height=1.20in,clip]{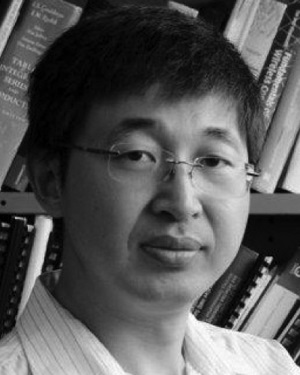}}]{Yonghui Li} (Fellow, IEEE) received the Ph.D. degree from the Beijing University of Aeronautics and Astronautics in November 2002.,Since 2003, he has been with the Centre of Excellence in Telecommunications, The University of Sydney, Australia. He is currently a Professor and the Director of the Wireless Engineering Laboratory, School of Electrical and Information Engineering, The University of Sydney. His current research interests include wireless communications, with a particular focus on MIMO, millimeter wave communications, machine-to-machine communications, coding techniques, and cooperative communications. He holds a number of patents granted and pending in these fields. He was a recipient of the Australian Queen Elizabeth II Fellowship in 2008 and the Australian Future Fellowship in 2012. He is an Editor of IEEE Transactions on Communications and IEEE Transactions on Vehicular Technology. He served as the Guest Editor for several IEEE journals, such as IEEE Journal on Selected Areas in Communications, IEEE Communications Magazine, IEEE Internet of Things Journal, and IEEE Access. He received the Best Paper Awards from IEEE International Conference on Communications (ICC) 2014, IEEE PIRMC 2017, and IEEE Wireless Days Conferences (WD) 2014.
\end{IEEEbiography}

\begin{IEEEbiography}[{\includegraphics[width=1.00in,height=1.20in,clip]{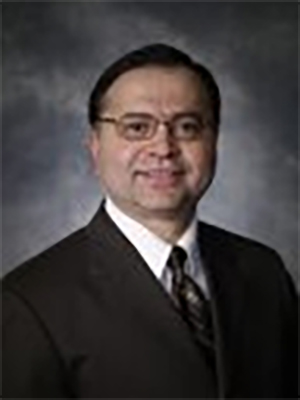}}] {Naofal Al-Dhahir} is Erik Jonsson Distinguished Professor and ECE Associate Head at UT-Dallas. He earned his PhD degree from Stanford University and was a principal member of technical staff at GE Research Center and AT\&T Shannon Laboratory from 1994 to 2003.  He is co-inventor of 43 issued patents, co-author of over 580 papers and co-recipient of 8 IEEE best paper awards. He is an IEEE Fellow, AAIA Fellow, received 2019 IEEE COMSOC SPCC technical recognition award, 2021 Qualcomm faculty award, and 2022 IEEE COMSOC RCC technical recognition award. He served as Editor-in-Chief of IEEE Transactions on Communications from Jan. 2016 to Dec. 2019.  He is a Fellow of the US National Academy of Inventors and a Member of the European Academy of Sciences and Arts.
\end{IEEEbiography}

\vfill

\end{document}